\documentclass[aps,prl,twocolumn,nofootinbib,superscriptaddress,showpacs,showkeys,10pt]{revtex4-1}
\usepackage[utf8]{inputenc}
\usepackage[british]{babel}
\usepackage{amsmath, amsxtra, amssymb}
\usepackage{graphicx, xspace, hyperref}
\usepackage{siunitx}
\usepackage{color}
\usepackage{textcomp}

\newcommand{\ket}[1]{\ensuremath{\lvert #1 \rangle}\xspace}

\newcommand{\expect}[3]{\ensuremath{\langle #1 \lvert #2 \rvert #3 \rangle}\xspace}

\renewcommand{\Re}{\operatorname{Re}}
\renewcommand{\Im}{\operatorname{Im}}

\begin{document}
\title{A subradiant optical mirror formed by a single structured atomic layer}
\author{Jun Rui}
\email[]{Jun.Rui@mpq.mpg.de}
\affiliation{Max-Planck-Institut f\"{u}r Quantenoptik, 85748 Garching, Germany}
\affiliation{Munich Center for Quantum Science and Technology (MCQST), 80799 Munich, Germany}

\author{David Wei}
\affiliation{Max-Planck-Institut f\"{u}r Quantenoptik, 85748 Garching, Germany}
\affiliation{Munich Center for Quantum Science and Technology (MCQST), 80799 Munich, Germany}

\author{Antonio Rubio-Abadal}
\affiliation{Max-Planck-Institut f\"{u}r Quantenoptik, 85748 Garching, Germany}
\affiliation{Munich Center for Quantum Science and Technology (MCQST), 80799 Munich, Germany}

\author{Simon Hollerith}
\affiliation{Max-Planck-Institut f\"{u}r Quantenoptik, 85748 Garching, Germany}
\affiliation{Munich Center for Quantum Science and Technology (MCQST), 80799 Munich, Germany}

\author{Johannes Zeiher}
\affiliation{Department of Physics, University of California, Berkeley, CA 94720, USA}

\author{Dan M. Stamper-Kurn}
\affiliation{Department of Physics, University of California, Berkeley, CA 94720, USA}

\author{Christian Gross}
\affiliation{Max-Planck-Institut f\"{u}r Quantenoptik, 85748 Garching, Germany}
\affiliation{Munich Center for Quantum Science and Technology (MCQST), 80799 Munich, Germany}
\affiliation{Physikalisches Institut, Eberhard Karls Universität Tübingen, 72076 Tübingen, Germany}

\author{Immanuel Bloch}
\affiliation{Max-Planck-Institut f\"{u}r Quantenoptik, 85748 Garching, Germany}
\affiliation{Munich Center for Quantum Science and Technology (MCQST), 80799 Munich, Germany}
\affiliation{Ludwig-Maximilians-Universität, 80799 Munich, Germany}

\date{\today}

%%%%%%%%%%%%%%%%%%%%%%%%%%%%%%%%%%%%%%%%%%%%%%%%%%%%%%%%%%%%
%                        Summary paragraph                            %
%%%%%%%%%%%%%%%%%%%%%%%%%%%%%%%%%%%%%%%%%%%%%%%%%%%%%%%%%%%%

%\begin{abstract}
%\end{abstract}

\maketitle

\textbf{Efficient and versatile interfaces for the interaction of light with matter are an essential cornerstone for quantum science \cite{chang:2018}. A fundamentally new avenue of controlling light-matter interactions has been recently proposed based on the rich interplay of photon-mediated dipole-dipole interactions in structured subwavelength arrays of quantum emitters \cite{porras:2008, jenkins:2012, jenkins:2013, gonzalez-tudela:2015, douglas:2015, facchinetti:2016, bettles:2016, shahmoon:2017, asenjo-garcia:2017}. Here we report on the direct observation of the cooperative subradiant response of a two-dimensional (2d) square array of atoms in an optical lattice. We observe a spectral narrowing of the collective atomic response well below the quantum-limited decay of individual atoms into free space. Through spatially resolved spectroscopic measurements, we show that the array acts as an efficient mirror formed by only a single monolayer of a few hundred atoms. By tuning the atom density in the array and by changing the ordering of the particles, we are able to control the cooperative response of the array and elucidate the interplay of spatial order and dipolar interactions for the collective properties of the ensemble. Bloch oscillations of the atoms out of the array enable us to dynamically control the reflectivity of the atomic mirror. Our work demonstrates efficient optical metamaterial engineering based on structured ensembles of atoms \cite{jenkins:2013, bettles:2016, shahmoon:2017} and paves the way towards the controlled many-body physics with light \cite{douglas:2015, gonzalez-tudela:2015, noh:2017} and novel light-matter interfaces at the single quantum level \cite{facchinetti:2016, asenjo-garcia:2017}}.

%%%%%%%%%%%%%%%%%%%%%%%%%%%%%%%%%%%%%%%%%%%%%%%%%%%%%%%%%%%%
%                 General Introduction                     %
%%%%%%%%%%%%%%%%%%%%%%%%%%%%%%%%%%%%%%%%%%%%%%%%%%%%%%%%%%%%

Cooperative optical effects in arrays of quantum emitters can be understood as resulting from the coherent scattering of photons between the emitters, or, equivalently, from optical dipole-dipole interactions \cite{svidzinsky:2010, shahmoon:2014}. Such interactions can result in drastic changes of the optical response of a suitably structured medium, a fact that is well known as Dicke super- or subradiance when the entire ensemble of atoms is confined to regions in space much smaller than the wavelength of light \cite{dicke:1954,gross:1982}. For extended spatially structured media, a much richer behaviour can emerge. For example by ordering only dozens of atoms into an array, one may obtain highly directional and tailored scattering properties based on strong cooperative light-matter interactions, with only the zeroth diffraction order dominant if the spacing of the array is smaller than the optical wavelength. The collective system features a modified optical resonance, due to an emergent cooperative Lamb shift, and super- or subradiant decay rates compared to those of single atoms \cite{porras:2008, jenkins:2013,facchinetti:2016, bettles:2016, shahmoon:2017, asenjo-garcia:2017, asenjo-garcia:2017a, chomaz:2012}. For subradiant modes, light can effectively be trapped and stored in the array, without the requirement for any external macroscopic (cavity) mirrors \cite{facchinetti:2016, asenjo-garcia:2017}. Exciting such subradiant modes, however, has proven to be challenging for experiments and so far limited to mesoscopic emitter arrays in the microwave regime \cite{jenkins:2017,vanLoo:2013,mirhosseini:2019}, systems of two ions \cite{devoe:1996} or small fractions of disordered atom clouds \cite{guerin:2016,solano:2017}. Inducing an overall collective subradiant response in the optical regime for an entire array of emitters has so far remained an outstanding achievement. Atomic dipoles held in free space in an optical lattice are of particular interest for this purpose, since they can be structured in a highly controllable fashion \cite{bloch:2008}, and both controlled and detected at the single-atom level \cite{bakr:2009,sherson:2010,weitenberg:2011}. Moreover, such trapped atoms are ideal quantum emitters at optical frequencies, whose dissipation is ascribed entirely to the coupling of the atoms to the electromagnetic vacuum. A collective subradiant response in such systems may be interpreted as a reduced coupling of the array to the quantum vacuum fields.

%%%%%%%%%%%%%%%%%%%%%%%%%%%%%%%%%%%%%%%%%%%%%%%%%%%%%%%%%%%%
%      Figure 1 - Introduction of the physical system      %
%%%%%%%%%%%%%%%%%%%%%%%%%%%%%%%%%%%%%%%%%%%%%%%%%%%%%%%%%%%%

\begin{figure*}[t!]
\centering
\includegraphics[width=1.9\columnwidth]{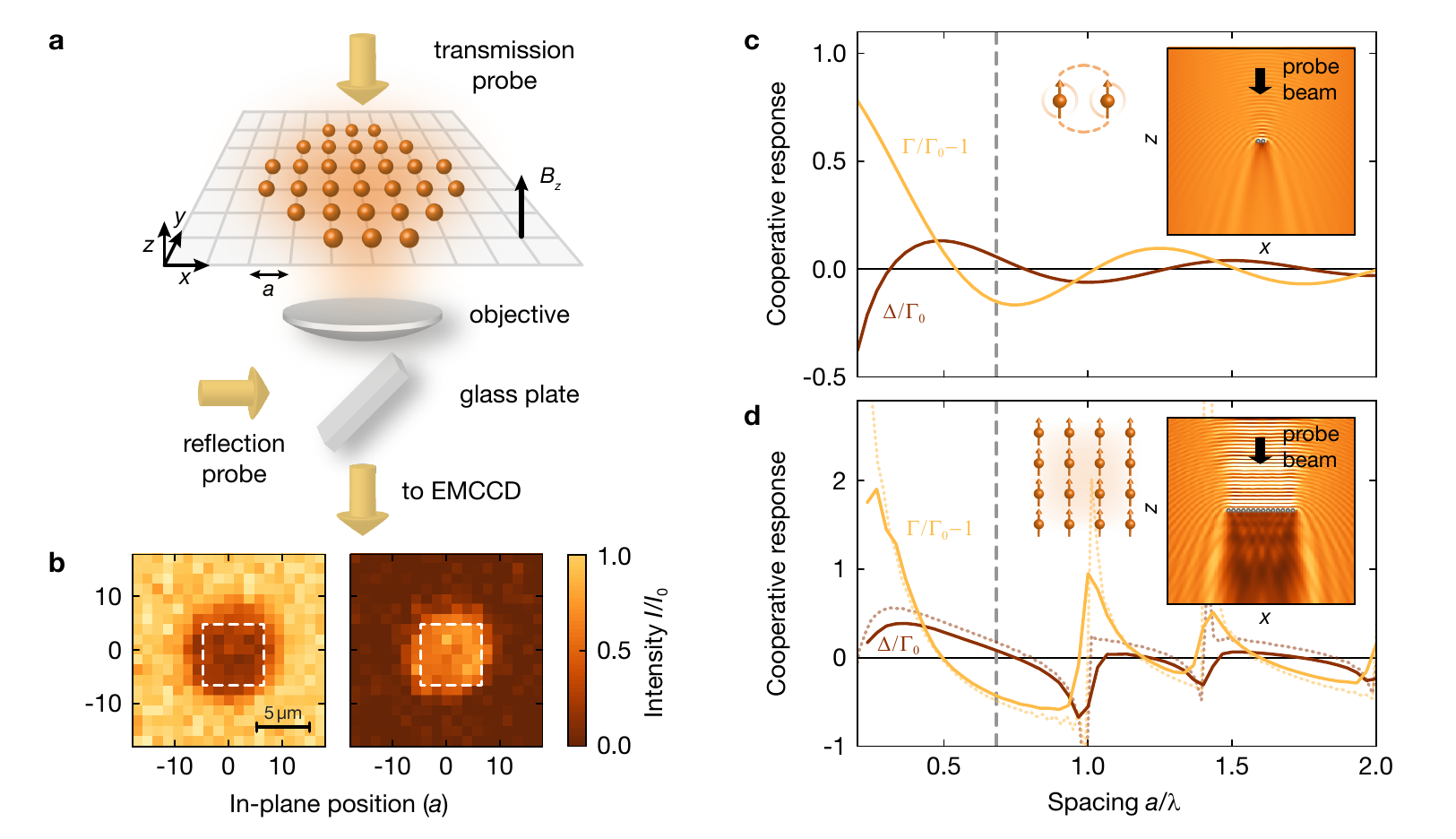}
  \caption{ \label{fig1}
  \textbf{Setup of the experiment and cooperative array response.}
(\textbf  {a}) We probe both transmission and reflection properties of the array using circularly polarized light fields propagating along the $\pm z$ direction, respectively. For the transmission measurement, both the probing and the residual transmitted field at the plane of the atoms can be collected and imaged onto the EMCCD, yielding the transmittance and absorptance of the array. For the reflection measurement, only the field reflected from the array is imaged. The nearly unity filled 2d array contains around 200 atoms in the experiment.
(\textbf  {b}) Average of the measured spatial profiles of the transmittance (left) and reflectance (right) from the atom array, with the region-of-interest used for analysis marked (dashed box). 
(\textbf  {c}) Simulated cooperative interactions between two circularly polarized point-like dipoles, including both the cooperative Lamb shift $\Delta/\Gamma_0$ as well as the cooperative correction to the decay rate $\Gamma/\Gamma_0-1$.
(\textbf  {d}) Simulated cooperative response of a 2d square lattice array unity filled with $14 \times 14$ atoms (solid line), including the finite wave packet spread of the ground state atoms. As a reference, the response of an infinitely large perfect 2d array of point-like dipoles is also shown (dotted line). The vertical dashed line corresponds to the lattice constant used in this work. 
(Inset) Cross-section plots of the optical fields at both sides of the pair or the 2d array of dipoles with a large uniform driving field. The forward scattered field interferes destructively with the driving field, reducing the optical transmission. The backward scattered field also interferes with the incoming field and forms standing waves. 
 }
\end{figure*}

The cooperative interactions inside an array of atomic dipoles can be intuitively understood by first considering the simpler case of only two dipoles, $\mathbf{d}_{i,j}$, placed at positions $\mathbf{r}_{i,j}$, respectively. Both dipoles have identical single atom polarizabilities $\alpha = -\alpha_0 \Gamma_0/(2\delta + i\Gamma_0)$, where $\alpha_0$ is the resonant polarizability, $\Gamma_0$ is the decay rate of the excited state and $\delta$ denotes the angular-frequency detuning from the optical resonance. When the two dipoles are close enough, each dipole $\mathbf{d}_i = \alpha (\mathbf{E}_0 (\mathbf{r}_i) + \mathbf{E}^{j}_\mathrm{sc} (\mathbf{r}_{i}))$ is driven by the external field $\mathbf{E}_0$ and the scattered field from the neighbouring atom $\mathbf{E}^{j}_\mathrm{sc} (\mathbf{r}_{i})$. Depending on the interatomic distance, these two fields interfere either constructively or destructively, leading to sub- or superradiant response of the composite system, as shown in Fig.~\ref{fig1}c.  Extending this picture to an ordered array, the scattered field $\mathbf{E}^{j}_\mathrm{sc} (\mathbf{r}_{i})$ is replaced by the fields emitted by all neighbouring dipoles. The coupling between the scattered fields and dipoles leads to a collective polarizability of the entire array, making it a new quantum object with a cooperative Lamb shift in the resonance frequency and modified decay rate from the collective excited state. For an infinitely large 2d square array, such collective states can be selectively created by choosing the incident angle of the driving field relative to the array \cite{asenjo-garcia:2017}. Changing the spacing $a$ of the array relative to the optical wavelength $\lambda$ leads again to oscillations between sub- and superradiant response, as shown in Fig.~\ref{fig1}d for a normal incident excitation field, where all atoms are driven in-phase relative to each other. In our case, the subwavelength array has a ratio of $a/\lambda=0.68$, allowing us to directly couple the subradiant mode with a normal incident drive field.

%%%%%%%%%%%%%%%%%%%%%%%%%%%%%%%%%%%%%%%%%%%%%%%%%%%%%%%%%%%%
%         Figure 2- Description of the experiment          %
%%%%%%%%%%%%%%%%%%%%%%%%%%%%%%%%%%%%%%%%%%%%%%%%%%%%%%%%%%%%

\begin{figure*}[htp]
\centering
\includegraphics[width=1.3\columnwidth]{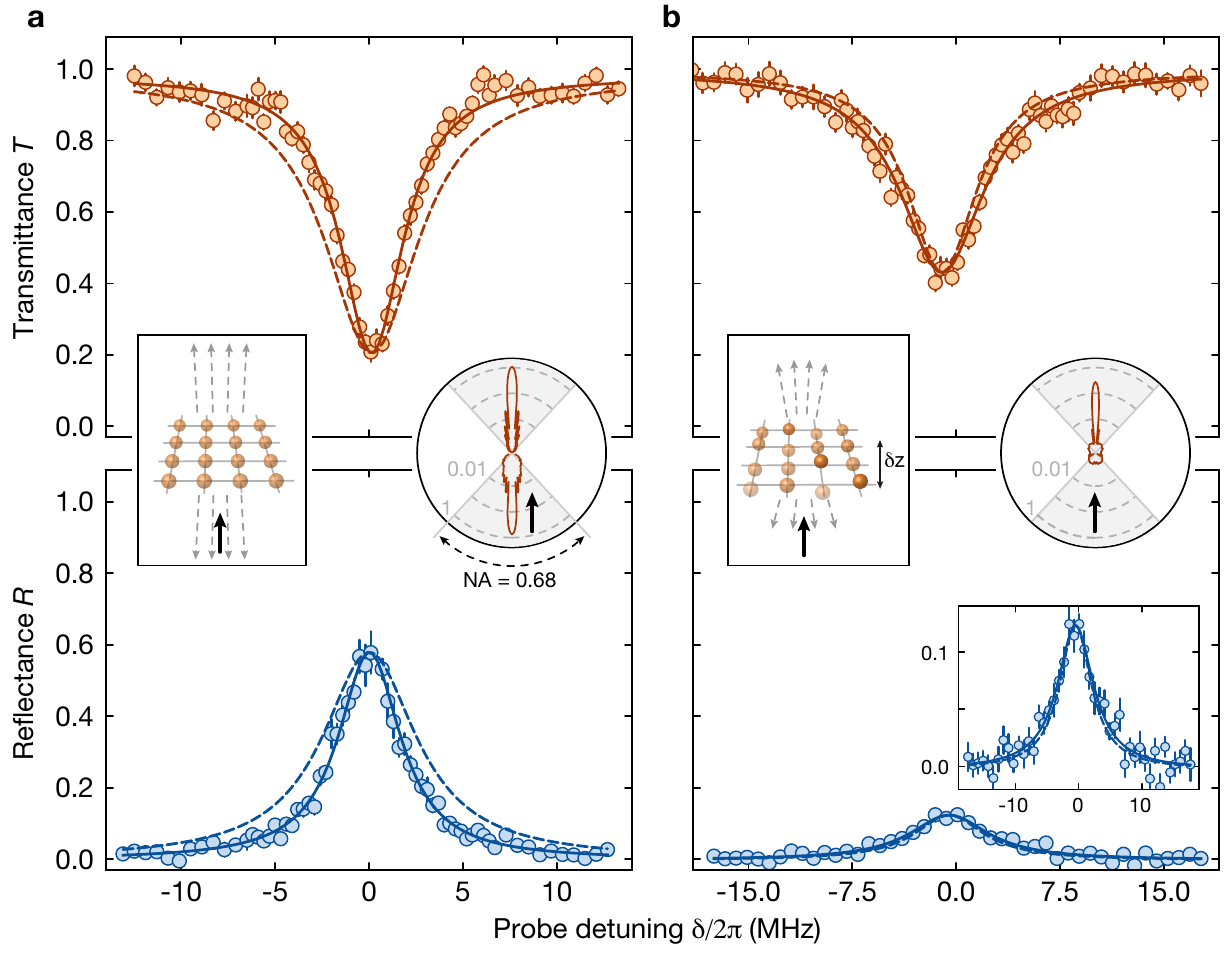}
  \caption{ \label{fig2}
  \textbf {Cooperative response for two different array geometries.} 
  (\textbf{a}) A nearly unity filled 2d array. The cooperative response is subradiant, with a linewidth of $\Gamma/2\pi = 4.09(11)$\,MHz and a transmittance of $T = 0.23(1)$ in the absorption spectrum, a linewidth of $\Gamma/2\pi =4.04(12)$\,MHz and a strong reflectance of $R = 0.58(3)$ in the reflection spectrum. 
  (\textbf{b}) A 3d array with vertical positions randomized. The linewidths become broadened to be $\Gamma/2\pi =7.38(28)$\,MHz and $\Gamma/2\pi =6.87(59)$\,MHz in the absorption and reflection spectra, with a transmittance of $T= 0.44(1)$ and suppressed reflectance of $R = 0.13(1)$, respectively.
  The dashed lines are reference spectra with the natural linewidth of single atoms. Each spectrum is an average of $15-30$ measurements. The insets in each figure sketch the ordering of the atoms and display the differential scattering cross-section obtained by numerical simulations, where the black arrow indicates the incident direction of the probe beam. Fits are based on Lorentzian line profiles and error bars in the spectra denote the standard error of the mean (s.e.m.).
 }
\end{figure*}

Our experiments started with a 2d array of $^{87}$Rb atoms loaded into a square optical lattice (lattice constant $a=532$\,nm), with nearly one atom per site \cite{sherson:2010}. Optical probing of the array was carried out on an isolated two-level transition in the $\text{D}_2$ manifold of $^{87}$Rb (see Methods). The collective response of the array was observed by recording either the forward or the backward scattering of a probing beam, as shown in Fig.~\ref{fig1}a and b. In the first configuration, the upper probe beam with field $\mathbf{E}_0$ was first scattered by the atom array, then the transmitted field $\mathbf{E}_{\text{trans}}$ was collected and imaged by an objective onto an electron-multiplying charge-coupled device (EMCCD), yielding the transmittance of the array $T=\vert \mathbf{E}_{\text{trans}} \vert ^2/\vert \mathbf{E}_{0}\vert ^2$. In the second configuration, the probe beam with field $\mathbf{E}_{1}$ was sent through the objective onto the atoms and the reflected light field $\mathbf{E}_{\text{refl}}$ imaged onto the camera, which gives the reflectance $R = \vert \mathbf{E}_{\text{refl}} \vert ^2/\vert \mathbf{E}_{1}\vert ^2$. With the high spatial resolution in the imaging setup, only the scattered light field within the bulk of the atom array is recorded for the following analysis. The array was probed well within the weak drive regime to avoid any saturation effects and subsequent photon-photon interactions in the array (see Methods).

%%%%%%%%%%%%%%%%%%%%%%%%%%%%%%%%%%%%%%%%%%%%%%%%%%%%%%
%          Figure 3 - Cooperative Response           %
%%%%%%%%%%%%%%%%%%%%%%%%%%%%%%%%%%%%%%%%%%%%%%%%%%%%%%

\begin{figure*}
\centering
\includegraphics[width=1.5\columnwidth]{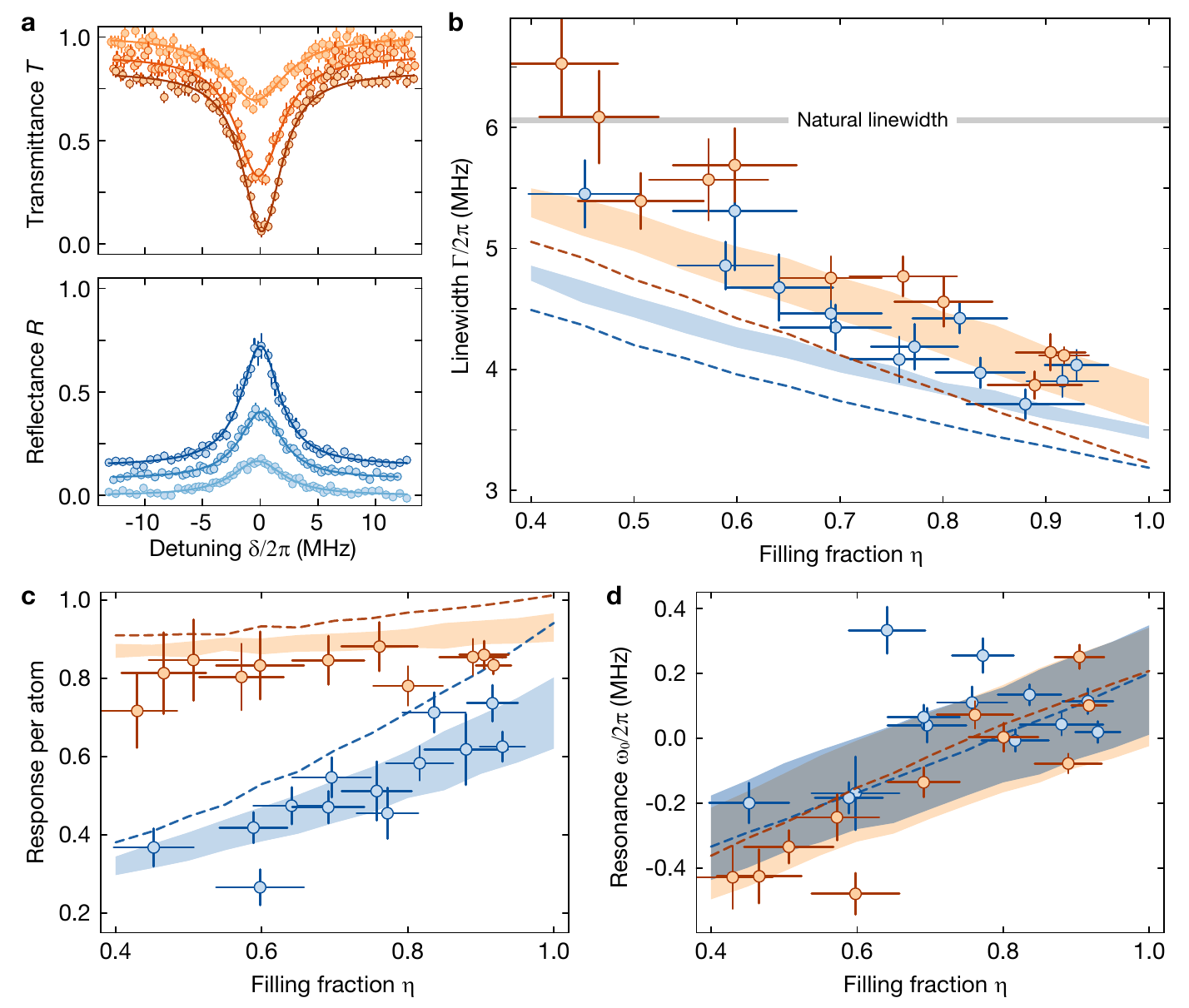}
  \caption{ \label{fig3}
\textbf  {Cooperative response versus filling fraction in the 2d array.} 
(\textbf{a}) Transmission and reflection spectra for three fillings of $\eta = 0.44, 0.69, 0.92$ (light to dark), with the vertical axis shifted accordingly.
(\textbf{b}) Fitted linewidths with different fillings. The optical response of the array changes from nearly non-interacting at low fillings to strongly subradiant at high fillings.
(\textbf{c}) Normalized optical response for each atom. While the absorptance roughly remains nearly constant, the reflectance shows significant dependence on the filling fraction, highlighting the cooperative contribution in the directional reflection.
(\textbf{d}) Dependence of the resonance frequency on the filling fraction, which reveals the cooperative Lamb shift. 
In all panels, color shaded regions correspond to the simulations with positional spreads between $0.054\,a$ and $0.14\,a$ along the $z$-axis and fixed spread of $0.054\,a$ along the $x$-/$y$- axes. The dashed lines correspond to simulations of ideal point-like scatters in a perfect array with similar size.
Error bars denote s.e.m. in the spectrum, and standard deviation (s.d.) of fitted parameters in others.
}
\end{figure*}

We first aim at identifying the spectral response for different spatial geometries of the atom cloud. The two geometries we compare correspond to (a) an almost unity filled 2d array, (b) a vertically disordered 3d array (see Fig.~\ref{fig2} and Methods). The transmittance and reflectance from these configurations was then probed as a function of the probe laser frequency. For the ordered array (a), we obtain transmission spectra showing a pronounced subradiant response with a cooperative linewidth of $\Gamma/\Gamma_0 = 0.68(2)$ and a residual transmittance of only $T=0.23(1)$ on resonance. The corresponding reflection measurement exhibits a reflectance of up to $R \simeq 0.58(3)$ and a linewidth of $\Gamma/\Gamma_0 = 0.66(2)$. The fact that the transmittance and reflectance do not add up to unity ($T+R<1$) originates from the finite collection angle of our objective, where fields scattered at large angles are not collected by the objective (see Supplementary Material). For configuration (b), we intentionally randomized the positions of the atoms along the vertical $z$-direction, while keeping their horizontal positions in the array fixed, resulting in the same vertical column density as the one of (a). Disrupting the vertical position order of the array leads to significant increase in transmittance and dramatic reduction in reflection, yielding $T=0.44(1)$ and $R=0.13(1)$. Moreover, the linewidth of the optical response is \emph{increased} beyond the natural linewidth, with $\Gamma/\Gamma_0=1.22(5)$ in transmission and $\Gamma/\Gamma_0=1.13(10)$ in reflection. In addition to that, we also studied the case of an in-plane disordered atomic ensemble and observed an increased linewidth, similar to the case of the vertically disordered array (see Supplementary Material). 

The strong difference in the cooperative linewidths for the above configurations confirms that the directly driven subradiant modes require the presence of a spatially ordered 2d array, as expected from the coupled dipole theory \cite{shahmoon:2017, asenjo-garcia:2017}, while the linewidth broadening in the random geometry can be attributed to an inhomogeneous broadening in both resonance frequencies and linewidths of the coupled modes (see Supplementary Material). The large difference in reflectance we observe for the two cases results from the fact that the 2d subwavelength array exhibits collectively enhanced scattering \cite{bettles:2016}, together with suppressed emission inside the atom plane. Such enhanced interaction is washed out between the emitters in the 3d array due to the randomized positions of the dipoles, leading to weaker induced dipole strengths and reduced collective interferences in the emissions. Besides that, additional position-dependent phases along the probe beam direction are imprinted onto the collective atomic state, which favours phase-matched emission along the same direction, while retro-reflection is suppressed \cite{porras:2008} (see differential scattering cross section in Fig.~\ref{fig2}).

%%%%%%%%%%%%%%%%%%%%%%%%%%%%%%%%%%%%%%%%%%%%%%%%%%%%%%
%      Figure 3     Change filling fraction          %
%%%%%%%%%%%%%%%%%%%%%%%%%%%%%%%%%%%%%%%%%%%%%%%%%%%%%%

To control the effective interaction strengths between the atoms, we varied the filling of the 2d array between $\eta = 0.4-0.9$ (see Methods) and characterized the change in the cooperative response. In Fig.~\ref{fig3}b, the fitted cooperative linewidths from the measured absorption and reflection spectra are both shown. For low fillings, the atom array shows a linewidth close to the one of isolated single atoms, whereas for increasing filling the cooperative response of the array becomes more and more subradiant, with the lowest observed linewidths of the array below $\Gamma < 2\pi \times 4$\,MHz, less than $65\%$ of the natural linewidth $\Gamma_0$. In Fig.~\ref{fig3}c, we compare the filling-normalized absorptance $A/\eta=(1-T)/\eta$ and reflectance $R/\eta$. The reflectance per atom increases notably with filling - a fact that can be directly attributed to the cooperative contribution within the directional reflection from the 2d array. The absorptance per atom, however, appears rather independent of the lattice filling. We attribute this to the stronger high-order diffractions at low fillings, due to the wavefront distortions in the transmitted field, leading to larger intensity extinction for a finite numerical aperture of the imaging setup and thus balancing the reduction in the cooperative response (see Supplementary Material). We also find a notable shift in the resonance frequency of the array, as a function of the filling. The frequency shift for increased fillings originates from the cooperative dipole-dipole interaction effect, and thus can be identified as the cooperative Lamb shift \cite{meir:2014} (see Fig.~\ref{fig3}d). All observed signals agree also quantitatively with our simulations of the classical coupled dipole equations. In the simulations, a position spread of each atom in the array on the order of the Gaussian ground state wave function size was included (see Supplementary Material).

%%%%%%%%%%%%%%%%%%%%%%%%%%%%%%%%%%%%%%%%%%%%%%%%%%%%%%%%%%%%
%                       Figure 4                           %
%%%%%%%%%%%%%%%%%%%%%%%%%%%%%%%%%%%%%%%%%%%%%%%%%%%%%%%%%%%%

\begin{figure}[t]
\centering
\includegraphics[width=1.0\columnwidth]{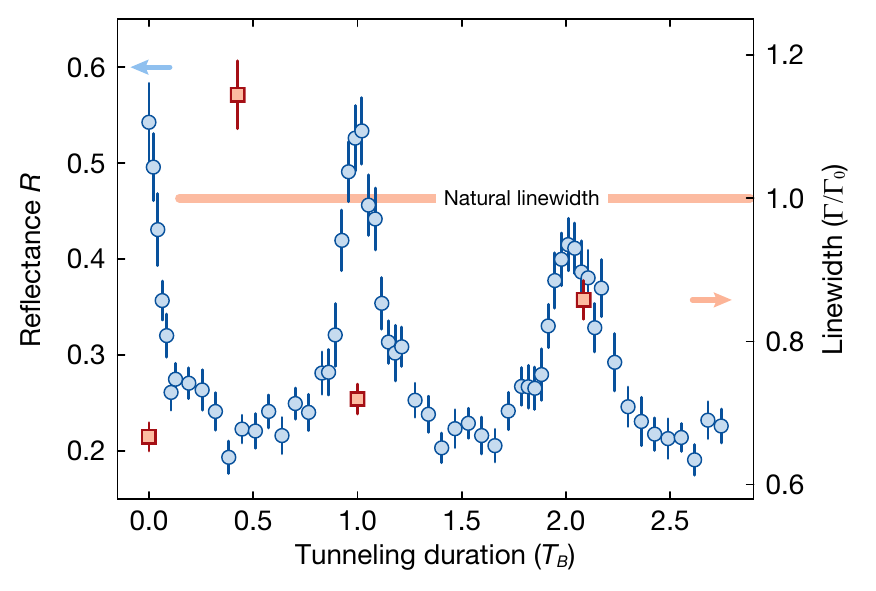}
  \caption{ \label{fig4}
\textbf  {Cooperative response under Bloch oscillation.} Starting from a nearly unity filled 2d array, the atomic wave packet periodically oscillates along the vertical direction. The reflectance (blue circles) is collectively enhanced once the atomic wave packet refocuses back to the original 2d plane. The lower height and reduced subradiance of the revival peaks results from the residual curvatures in the potential gradients. The measured linewidth (red squares) clearly shows the dynamical recovering of subradiant response. Error bars denote s.e.m.\ for reflectance, and s.d.\ for fitted linewidth.
 }
\end{figure}

Utilizing the high degree of control over the atoms in the optical lattice, one can dynamically dissolve and restore the 2d ordering of the atoms and thereby dynamically switch the subradiant response through quantum dynamics. For this, we let the atoms perform Bloch oscillations inside the vertical lattice along the $z$-direction under the action of a small potential gradient, giving rise to an energy difference of $\Delta_{z}$ between neighbouring lattice sites. The initially localized atoms thereby periodically spread out in the vertical direction and refocus to the original position at integer multiples of the Bloch period $T_B = h /\Delta_z \simeq 4.7$\,ms for the experimental parameters (see Methods) with $h$ the Planck constant \cite{preiss:2015}. This results in a periodic change of the geometry from an ordered array in 2d to a disordered array (along the $z$-direction) in 3d, with an estimated maximum half-width of $\delta z \approx 2.5\,a$. The measured reflectance versus the tunnelling time in the tilted lattice is shown in Fig.~\ref{fig4}. In the subsequent dynamics two reflection peaks appear at the times of optimal re-focusing. The reflection becomes significantly suppressed once the atoms slightly tunnel out of the single plane. The measured linewidths at the two reflection peaks are both subradiant. In contrast, the measured linewidths at the first half period, where the atoms maximally disperse out of the plane, is broadened to a value above the natural linewidth.

%%%%%%%%%%%%%%%%%%%%%%%%%%%%%%%%%%%%%%%%%%%%%%%%%%%%%%%%%%%%
%                        Figure 5                          %
%%%%%%%%%%%%%%%%%%%%%%%%%%%%%%%%%%%%%%%%%%%%%%%%%%%%%%%%%%%%

\begin{figure*}[thp]
\centering
\includegraphics[width=0.75\textwidth]{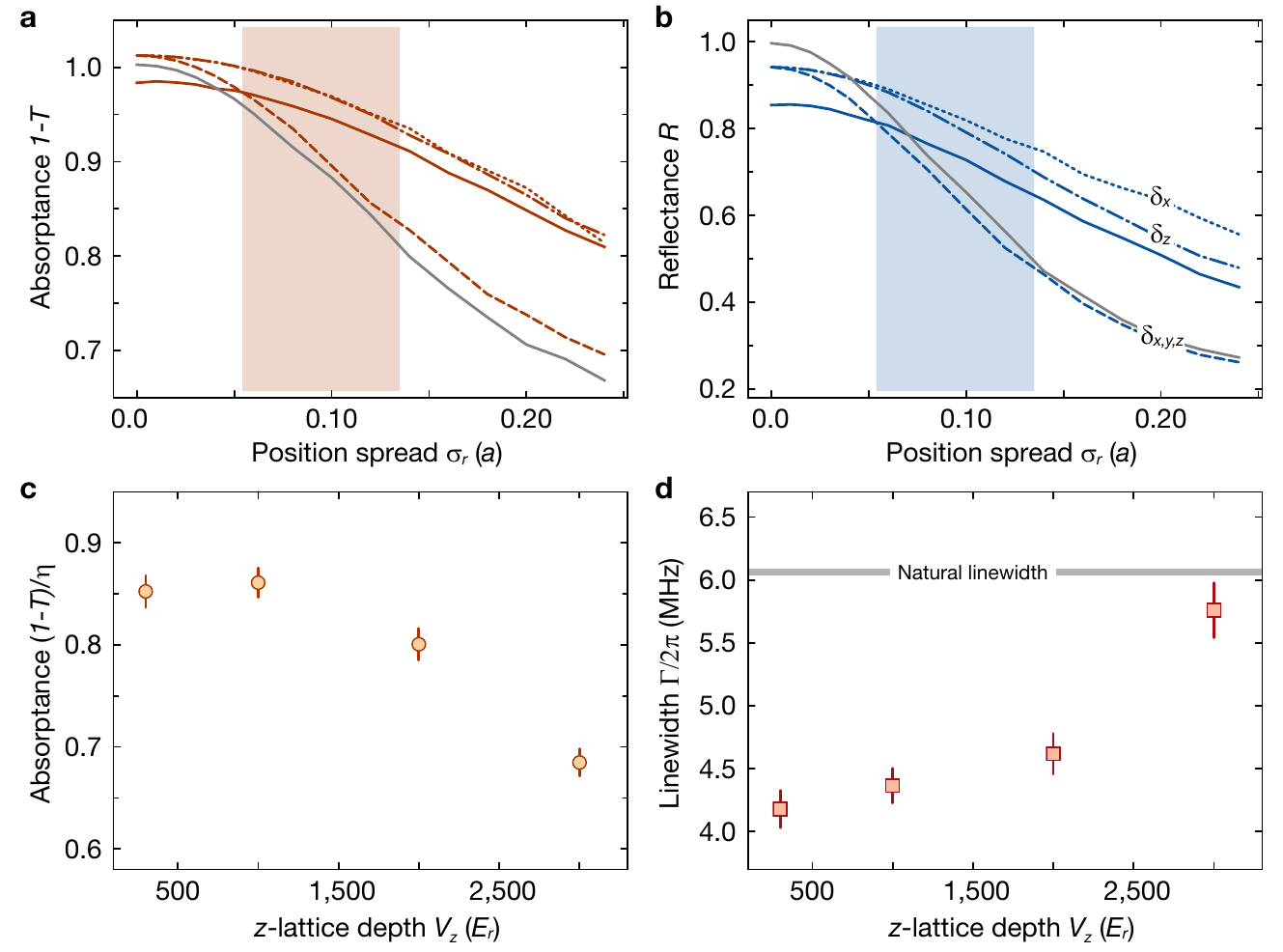}
\caption{\label{fig5}
\textbf{Limitations to the cooperative response of the 2d array.} 
(\textbf{a}, \textbf{b}), Simulated absorptance and reflectance with different positional spreads in the array with similar size as in the experiment. The dotted (dash-dotted) lines represent positional spread only along the $x$- ($z$-) axis. The dashed line represents the same spread along all three axes. The solid lines (red and blue) represent variable spread along the $z$-axis, but with a fixed spread of $0.054\,a$ along the $x$-/$y$- axis, corresponding to the ground state spread of a $300\,E_r$ lattice depth, with $E_r$ denoting the recoil energy of a single atom. Color shaded area together with the solid line corresponds to the parameter region used in the simulation in Fig.~\ref{fig3}. The imperfect reflectance ($R\simeq0.95$) at the zero spread limit is a result of the edge effect due to the large beam size used in the simulation (beam waist $\omega_0=56\,a$). As a comparison, the grey solid line represents the simulation with equal spread along all directions and a probe beam with size smaller than that of the 2d array ($\omega_0=6\,a$), with $R\simeq1$ at the zero spread limit. 
(\textbf{c}, \textbf{d}) Cooperative response measured under different vertical lattice depths for nearly unity filled 2d array. While the filling normalized absorptance significantly decreases in deep lattices, the measured linewidth signals show reduced subradiance for increasing lattice depths. The depths of the horizontal lattices were fixed at $V_{x,y} \simeq 1000\,E_r$ during the measurement. 
Error bars denote s.d. of the fits. 
}
\end{figure*}

Despite the strongly subradiant response of the ordered 2d array, deviations from an ideal perfect array of point-like emitters persist in the experiment, which can be attributed to the finite spread of the atomic wave function in each lattice site. The spread essentially disorders the dipoles from positions in a perfect array along all directions during the photon scatterings, leading to weakened cooperative interactions and reduced reflectivity of the atom mirror, as shown in Fig.~\ref{fig5}b. Naively, one would expect that going to deeper lattices could improve the reflectivity close to $R\approx 1$, and reduce the linewidth to $\Gamma/\Gamma_0 = 0.56 (0.51)$ as in the case of a perfect array of similar (infinite) size by reducing the zero-point motion in extent. However in the experiment, we observe the opposite behaviour, as shown in Fig.~\ref{fig5}c,d. When the potential depth of the vertical lattice is increased, the measured linewidth continuously increases as well. This effect can mainly be explained through a motional spreading experienced by atoms in the anti-trapped excited $5^{2}P_{3/2}$ state in the lattice potential, which leads to enhanced heating after decaying back to the ground state together with position-dependent shifts of the transition frequency inside each lattice site. Both effects become stronger for deeper lattices and therefore lead to the breakdown of the cooperative response of the array. A quantitative interpretation of the effects would require a full quantum treatment of the coupling between internal and motional degrees of freedom of the atoms, which is lacking so far. We note that a system where a trapping laser on a `magic' transition is used, e.g. in optical lattice clocks \cite{ye:2008}, should not exhibit this heating effect. Based on our numerical simulations, we are able to extrapolate the experimental reflectance of the array to $R>0.8$ in the limit where only a single or few photons are scattered and heating effects are thus negligible (see Fig.~\ref{fig5}b). Larger arrays or probe beams smaller than the array size would reduce detrimental edge effects present in the current experiment and should allow one to further increase the reflectance to $R>0.9$ (see also Supplementary Material).

%%%%%%%%%%%%%%%%%%%%%%%%%%%%%%%%%%%%%%%%%%%%%%%%%%%%%%
%                         extended figure            %
%%%%%%%%%%%%%%%%%%%%%%%%%%%%%%%%%%%%%%%%%%%%%%%%%%%%%%

%%%%%%%%%%%%%%%%%%%%%%%%%%%%%%%%%%%%%%%%%%%%%%%%%%%%%%
%                     Outlook                        %
%%%%%%%%%%%%%%%%%%%%%%%%%%%%%%%%%%%%%%%%%%%%%%%%%%%%%%

Our local resolution of the atomic array provides an ideal setting to explore collective optical excitation transport, for example, in topologically protected edge states~\cite{perczel:2017,bettles:2017}. Furthermore, subradiant states could be employed to design new light-matter interfaces without the need for any external cavities or dielectric media \cite{scully:2015, facchinetti:2016,asenjo-garcia:2017}. One could even envisage cavities formed by two atomic mirror arrays or scenarios where an external mirror is combined with an atomic mirror array \cite{guimond:2019, cernotik:2019}, to realize novel cavity QED systems. Since the atomic arrays form the lightest possible mirrors, they also open up unique opportunities for exploring the rich mechanical effects of light interacting with the mirror array \cite{shahmoon:2019,shahmoon:2018}. Adding Rydberg impurities could allow one to form novel quantum metasurface states, with the mirror being in a superposition state of low and high reflectance \cite{bekenstein:2019}. Finally, when driving such an array close to saturation, a full quantum mechanical many-body treatment of the photons in the array is needed, opening the path to interacting many-body physics with optical photons \cite{olmos:2013, asenjo-garcia:2017, Henriet2019, zhang:2019, bettles:2019}. The flexible platform of atoms in optical lattices allows to explore all these directions in the future.

\bigskip

\begin{acknowledgments}
\textbf{Acknowledgements:}
  We thank J. Ruostekoski, E. Shahmoon, J. I. Cirac, R. Bettles, R. Bekenstein, S. Yelin and M. Zwierlein for valuable discussions.
  We acknowledge funding by the Max Planck Society (MPG), the European Union (PASQuanS Grant No. 817482) and the Deutsche Forschungsgemeinschaft (DFG, German Research Foundation) under Germany’s Excellence Strategy – EXC-2111 – 390814868. J.R. acknowledges funding from the Max Planck Harvard Research Center for Quantum Optics. J.Z acknowledges support through a Feodor Lynen Fellowship by the Humboldt Foundation. D.\,S.-K. acknowledges support through a Carl Friedrich von Siemens Research Award of the Alexander von Humboldt Foundation.
\end{acknowledgments}

%\clearpage 

%%%%%%%%%%%%%%%%%%%%%%%%%%%%%%%%%%%%%%%%%%%%%%%%%%%%%%
%                      Methods                       %
%%%%%%%%%%%%%%%%%%%%%%%%%%%%%%%%%%%%%%%%%%%%%%%%%%%%%%

\section{Methods}

\textbf{Preparation of the atom array.}
We started the experiment by selecting a single layer of atoms out of a Bose-Einstein condensate loaded into a vertical optical lattice. After that, we ramped up two horizontal lattices to create a Mott-insulating state, and prepared a 2d array of $^{87}$Rb atoms with a maximum filling of $\eta \simeq 0.92$ per lattice site on $\simeq 200$ lattice sites. All of the atoms were prepared in the motional ground state of the 3d optical lattices, and initialized in the ground Zeeman sublevel of $|F=1, m_F=-1\rangle$, then transferred into the final $|F=2, m_F=-2\rangle$ state with a microwave (MW) sweep. Here $F$ and $m_F$ are the hyperfine and magnetic quantum numbers, respectively. A small stabilized magnetic field of $3.3$\,G was applied perpendicular to the atomic plane to isolate a two-level response with the $\vert F=2, m_F=-2\rangle \leftrightarrow \vert F'=3, m'_F = -3\rangle$ transition in the $D_2$ manifold of $^{87}$Rb, which has a natural linewidth of $\Gamma_0 = 2\pi \times 6.06$\,MHz. The probe light was $\sigma^-$-polarized and frequency-locked to an ultra low expansion cavity with a final laser linewidth of $<50$ kHz. Before optical probing, all three lattices were adiabatically ramped up to a depth of $300\,E_r$ in order to pin the atoms, with $E_r$ denoting the recoil energy of a single atom $E_r=h^2/(8 m a^2)$, $h$ the Planck constant and $m$ the mass of a $^{87}$Rb atom. With the given lattice depth, the RMS spread of the atom position is $\sigma_0=0.054\,a$ in the motional ground state of each lattice site.

To control the filling of the 2d array, we used MW driven Rabi oscillations to partially transfer the atoms to the $|F=2, m_F=-2\rangle$ state from the initial $|F=1,  m_F=-1\rangle$ state, followed by a resonant optical push-out of the atoms in the $F=2$ state, to reduce the filling in a controlled way. Before the optical probe, atoms left in the $F=1$ state were MW swept back into the $|F=2, m_F=-2\rangle$ state. The resulting atom clouds with reduced filling were independently characterized by single-site resolved fluorescence imaging \cite{sherson:2010}.

\textbf{Detection scheme and probe intensity.}
The optical field at the plane of the 2d atom array was imaged by a high resolution objective with a numerical aperture (NA) of 0.68 and a diffraction limit of about $700$\,nm. In order to increase our signal-to-noise ratio for the small probe photon numbers, we electronically binned the CCD readout in $8 \times 8$ pixel clusters, corresponding to an area of $\simeq 3.5$ lattice sites in the atom plane. The transmission efficiency through the entire imaging path is estimated to be $61\%$, and the quantum efficiency of the camera is $80\%$. For the absorption measurements, we used a fluence of about $20$ photons per lattice site in the probe beam within a duration of $3$ ms. For the reflection measurements, a small fraction ($\sim 4\%$) of the reflection probe beam is reflected by a glass plate and focused onto the atom plane after passing through the objective, with an estimated fluence of approximately $27$ or $50$ photons per lattice site within a duration of $5$\,ms or $10$\,ms (see calibration in Supplementary Material). In the experiment, we could not observe a significant difference in the reflectance and linewidth between the two reflection intensities. We confirmed that the $\sigma^-$ polarization purity of both probe beams is larger than $98\%$. The sizes of both probe beams are significantly larger than the size of the atom array.

\textbf{Control of spatial geometries.}
To prepare the vertically disordered array, we started with the nearly unity filled 2d array at $V_{x, y} \simeq 40\, E_r$ for the horizontal lattices, and $V_z \simeq 13\, E_r$ for the vertical lattice. Then we suddenly switched off the $z$-lattice to let the atoms freely expand during $1$\,ms along $z$. Subsequently, we rapidly ramped the vertical lattice depth up to $\simeq 16\, E_r$ to pin down the positions of the atoms. At the end, all lattices were slowly ramped up to $300\,E_r$ for optical probing. We estimate that the 3d disordered array finally occupies an RMS size of 10 lattice sites along the vertical direction after the free expansion.

For the Bloch oscillations, we also started with the same 2d array at $40\,E_r$ in the horizontal lattices and $16\, E_r$ in the $z$-lattice. Then the depth of the $z$-lattice was suddenly reduced to $5~E_r$ to start the periodic evolution. After a variable tunnelling duration, the vertical lattice depth was first quickly ramped up to $13~E_r$, then all three lattice depths were slowly ramped up to $300~E_r$ for optical probing. The gradient field used to drive the Bloch oscillation was provided by the vertical potential gradient in the horizontal lattice beams together with gravity gradient. The maximum half-width during the Bloch oscillation is estimated to be $\delta z = 4Ja/\Delta_z=2.5\,a$, with $J$ the tunnelling rate in the vertical lattice.

\textbf{Simulation of electromagnetic response.}
The simulation results were obtained by directly solving the $\sigma^-$-polarization-projected coupled dipoles equations, $\mathbf{d}_l = \alpha_l (\Delta) \hat{P}_{\sigma^-} ( \mathbf{E}_0 (\mathbf{r}_l) + \sum_{j \neq l} \hat{G} (\mathbf{r}_l - \mathbf{r}_j) \mathbf{d}_j )$, with $\hat{G}(\mathbf{r})$ the free space dyadic Green function, yielding the individual dipole moments $\mathbf{d}_l$ for a given driving field $\mathbf{E}_0$. By calculating the far field of the dipole emission pattern $\mathbf{E}_\mathrm{sc} (\mathbf{r}) = \sum_j \hat{G} (\mathbf{r} - \mathbf{r}_j) \mathbf{d}_j$ and of the probe beam within the collection angle of the objective, we obtain the reflectance and absorptance. The linewidths, resonance detunings and maximum response amplitudes are determined by performing the calculations for different driving field detunings $\Delta$ (and thus different polarizabilities $\alpha_l$) and fitting a Lorentzian function. The finite positional spread is modelled by randomly sampling the dipole positions according to the spatial density distribution and averaging the far field intensities.

\bibliography{Atommirror_revtex}

%%%%%%%%%%%%%%%%%%%%%%%%%%%%%%%%%%%%%%%%%%%%%%%%%%%%%%
%                         supplement            %
%%%%%%%%%%%%%%%%%%%%%%%%%%%%%%%%%%%%%%%%%%%%%%%%%%%%%%

\clearpage 

\setcounter{equation}{0}
\setcounter{figure}{0}
\renewcommand{\theequation}{S\arabic{equation}}
\renewcommand{\thefigure}{S\arabic{figure}}

\section{Supplementary Information}

%\tableofcontents

%%%%%%%%%%%%%%%%%%%%%%%%%%%%%%%%%%%%%%%%%%%%%%%%%%%%%%%%%%%%%%%%%%%%%%%
% Simulation                                                          %
%%%%%%%%%%%%%%%%%%%%%%%%%%%%%%%%%%%%%%%%%%%%%%%%%%%%%%%%%%%%%%%%%%%%%%%
\section{Simulation}

Here we describe our theoretical model and present more detailed simulation results corresponding to the experimental measurements.

\subsection{Coupled dipoles model}

Since atomic two-level systems (with states $\ket{g}, \ket{e^p}$) behave identical to classical dipoles in the weakly driven limit, the atomic system can be modelled with the coupled dipole equations (as outlined e.g. in \cite{chomaz:2012, jenkins:2012}), which are equations of motion for the atomic dipole moments $\mathbf{d}_l$ and are given by
\begin{align}
  \dot{\mathbf{d}}_l = &\sum_p \left( i (\delta - \delta_l^p) - \frac{\Gamma_{0, l}^p}{2} \right) \hat{P}_p \mathbf{d}_l + \notag \\
 &+ \frac{i}{\hbar} \sum_p |{\wp}_p|^2 \hat{P}_p \left( \mathbf{E}_0 (\mathbf{r}_l) + \sum_{j \neq l} \hat{G} (\mathbf{r}_{lj}) \mathbf{d}_j \right),
\end{align}
where $j, l$ denote the $j$th/$l$th atom, $\mathbf{r}_{lj} = \mathbf{r}_l - \mathbf{r}_j$ their distance, and $\delta$ the detuning of the drive. $\delta_l^p$ and $\Gamma_{0, l}^p = \alpha_0 |\wp_l^p|^2 / \hbar$ are the local resonance detuning and the natural decay rate of the respective atom, $\wp_l^p = \expect{g_l}{\hat{d}^p}{e_l^p}$ the dipole matrix element, $\alpha_0 = 6 \pi \varepsilon_0 / k^3$ the resonant polarizability with wavenumber $k$, and $\hat{P}_p$ is the projector onto polarization $p \in \{ \sigma^\pm, \pi \}$. The dyadic Green's function
\begin{align}
  \hat{G} (\mathbf{r}) = \frac{k^2 e^{i k r}}{4 \pi \varepsilon_0 r} & \left[ \left( 1 + \frac{i}{k r} - \frac{1}{(k r)^2} \right) \mathbb{I} \right. + \notag \\
  & \enskip \left. \left( -1 - \frac{3}{k r} + \frac{3}{(k r)^2} \right) \hat{\mathbf{r}} \otimes \hat{\mathbf{r}} \right],
\end{align}
with radial unit vector $\hat{\mathbf{r}}$, propagates the field emitted by a single dipole, yielding the total field emitted by all dipoles, $\mathbf{E}_\mathrm{sc} (\mathbf{r}) = \sum_j \hat{G} (\mathbf{r} - \mathbf{r}_j) \mathbf{d}_j$.

In the quickly reached steady state $\dot{\mathbf{d}}_l = 0$, the equations of motion simplify to the coupled linear relations
\begin{equation}
  \frac{1}{\alpha_l^p (\delta)} \mathbf{d}_l^p = \hat{P}_p \left( \mathbf{E}_0 (\mathbf{r}_l) + \sum_{j \neq l} \hat{G} (\mathbf{r}_{lj}) \mathbf{d}_j \right),
  \label{eq:coupled_dipoles}
\end{equation}
with atomic polarizability $\alpha_l^p (\delta) = -\alpha_0 (\Gamma_{0, l}^p / 2) / ((\delta - \delta_l^p) + i (\Gamma_{0, l}^p / 2))$. The dipole moments $\mathbf{d}_l$ at certain drive detunings $\delta$ can then be self-consistently solved for.

\subsubsection{Cooperative eigenmodes}

If the atoms have identical responses, $\delta_l^p = 0, \Gamma_{0, l}^p = \Gamma_0, \alpha_l^p = \alpha$, the coupled dipoles equations (\ref{eq:coupled_dipoles}), projected to a polarization subspace $P$, reduce to $\hat{P}_P \mathbf{E}_0 (\mathbf{r}_l) = \hat{P}_P (1 / \alpha (\delta) + \sum_{j \neq l} \hat{G} (\mathbf{r}_{lj})) \mathbf{d}_j (\delta)$. As the frequency dependent part of the matrix is constant and diagonal, the behaviour of the cooperative eigenmodes is fully determined by the eigensystem of the Green's function $\hat{G}$. Its eigenvalues $\mu_q$ are related to the resonance detuning and linewidth as $\Delta_q = -\alpha_0 \Re [\mu_q] \Gamma_0 / 2$ and $\Gamma_q = (1 + \alpha_0 \Im [\mu_q]) \Gamma_0$, respectively.

The driving field $\mathbf{E}_0$ can then be treated as exciting certain cooperative modes $\mathbf{m}_q$ with expansion coefficient $c_q$, $\hat{P}_P \mathbf{E}_0 = \sum_q c_q \mathbf{m}_q$. The eigenmode expansion of the dipole moments $b_q (\delta)$ is obtained by dividing through the respective eigenvalues, $b_q (\delta) = c_q / (1 / \alpha (\delta) + \mu_q)$. To quantify the contribution of each dipole mode to the response, we define the dipole mode amplitudes as $|b_q (\delta) \mathbf{m}_q|^2$.

\subsubsection{Motional state}

The atoms are trapped in a deep three-dimensional optical square lattice $V (\mathbf{r}) = \sum_i V_i \sin^2 (\pi r_i / a)$ with spacing $a$. Therefore, the dipoles on each lattice site have a Gaussian spatial density distribution with a standard deviation of $\sigma_i$.

As the scattering process with a rate on the order of $\Gamma_0$ happens faster than the in-trap dynamics characterized by the trap frequency $\omega_\mathrm{tr} \ll \Gamma_0$, the atoms appear motionally frozen during the scattering process. To account for the finite positional spread, the dipole positions are randomly sampled according to the spatial probability distribution, over which the electromagnetic response is averaged.

\subsubsection{Energy flux}

The scattering properties of the dipole array due to a given driving field can be characterized through the differential cross section
\begin{align}
  \frac{d \sigma}{d \Omega} = & \frac{1}{\left| \mathbf{E}_0 \right|^2} \left( r^2 \left| \mathbf{E} (\mathbf{r}) \right|^2 - \left| \mathbf{r} \cdot \mathbf{E} (\mathbf{r}) \right|^2 \right) \notag \\
  \equiv & \frac{d \sigma_0}{d \Omega} + \frac{d \sigma_\mathrm{sc}}{d \Omega} + \frac{d \sigma_\mathrm{intf}}{d \Omega},
\end{align}
for which the expansion results from the sum of driving field $\left| \mathbf{E}_0 \right|^2$, scattered field $\left| \mathbf{E}_\mathrm{sc} \right|^2$ and their mutual interference field $\mathbf{E}_0^* \cdot \mathbf{E}_\mathrm{sc}$.
Assuming a Gaussian drive field with waist $w_0$, this results in
\begin{align}
  \frac{d \sigma_\mathrm{sc}}{d \Omega} = \frac{2 c \varepsilon_0}{I_0} \left( \frac{k^2}{4 \pi \varepsilon_0} \right)^2 & \left( \left| \sum_i \mathbf{d}_i e^{-i k \hat{\mathbf{r}} \cdot \mathbf{r}_i} \right|^2 \right. - \notag \\
  & - \left. \left| \sum_i \hat{\mathbf{r}} \cdot \mathbf{d}_i e^{-i k \hat{\mathbf{r}} \cdot \mathbf{r}_i} \right|^2 \right),
\end{align}
and in forward direction $(z > 0)$
\begin{equation}
  \frac{d \sigma_\mathrm{intf}}{d \Omega} = -\frac{k^3 w_0^2 e^{-(k w_0 \theta / 2)^2}}{4 \pi \varepsilon_0 |\mathbf{E}_0|^2} \sum_i \Im \left[ \mathbf{E}_0^* (\mathbf{r}_i) \cdot \mathbf{d}_i e^{-i k \hat{\mathbf{r}} \cdot \mathbf{r}_i} \right].
\end{equation}

In order to estimate the experimental observables, the finite light collection solid angle of the objective, characterized by the numerical aperture $\mathrm{NA} = \sin \theta_\mathrm{obj}$, has to be taken into account by numerically integrating
\begin{equation}
    \sigma^\mathrm{NA} = \int_{\sin \theta < \mathrm{NA}} \frac{d \sigma}{d \Omega} d \Omega.
\end{equation}

In the reflection measurements only the scattered light is detected, such that the reflectance becomes $R = \sigma_\mathrm{sc}^\mathrm{NA} / \sigma_\mathrm{in}$, where $\sigma_\mathrm{in}$ denotes a normalization cross section. In the transmission measurements the superposition field with the drive field is probed, such that the absorptance (non-transmittance) is defined to be $A = 1 - T = -(\sigma_\mathrm{sc}^\mathrm{NA} + \sigma_\mathrm{intf}^\mathrm{NA}) / \sigma_\mathrm{in}$.

Since the experimental setup is in an imaging configuration, the location of the cloud is resolved, so the normalization power $P_\mathrm{in} = I_0 \sigma_\mathrm{in}$ is not the full incident power $ P_0 \neq P_\mathrm{in}$. Assuming that any reflected/absorbed power originates from the cloud position, we can instead normalize by the power incident on the atoms $P_\mathrm{in} = \int_\mathrm{array} I (\mathbf{r}) d^2 r$.

% ++++++++++++++++++++++++++++++++++++++++

\subsection{Numerical evaluation}

For the numerical simulation, the reflectance and absorptance spectra of a $14 \times 14$ site square lattice with a subwavelength spacing of $a / \lambda = 0.68$ are calculated, with a driving beam size much larger than the atomic cloud size. These are fitted by a Lorentzian function, from which the resonance detuning, linewidth and on-resonance reflectance/absorptance are obtained.

\subsubsection{Local resonance detuning}

The optical lattice used in the experiment is in a configuration, in which the excited state atoms $\ket{e}$ are anti-trapped with respect to the ground state $\ket{g}$, resulting in the excited state potential $V_e (\mathbf{r}) \approx \hbar \omega_0 - 1.5 V_g (\mathbf{r})$. This however means that the resonance frequency $\omega_0$ of the transition becomes position dependent.

For a finite positional spread of the dipoles, the randomly drawn positions give rise to a shift in the atomic resonance, which is modelled as a local detuning $\delta_l^p$-term in the coupled dipole equations (\ref{eq:coupled_dipoles}).

\subsubsection{Anisotropic polarizability}

The atoms are subjected to a magnetic field of $B = 3.3\,\mathrm{G}$ along the optical axis, are prepared in the $\ket{F = 2, m_F = -2}$ ground state and are probed on the $\mathrm{D}_2$ transition with a $\sigma^-$-polarized beam on their closed transition to $\ket{F' = 3, m_F = -3}$. The Zeeman shift detunes the $\sigma^+$ transition by $\delta_{\sigma^+} \simeq \Gamma_0$, and thus out of resonance.

The $\pi$ transition is not as far detuned, but can also not be directly driven by the probe beam due to the mutual orientation. The rescattered field might have $\pi$ polarization but is typically significantly weaker than the probe beam. Additionally, for a two-dimensional geometry, the dipole emission pattern prohibits $\pi$ polarization within the plane. Therefore we assume that we only operate within the $\sigma^-$ polarization subspace.

\subsubsection{Trap positions}

The random positional sampling described above only samples the probability density within a trap. The trap positions themselves (depending on the measurement configurations) are determined before calculating the dipole response. To simulate:

\begin{itemize}
  \item Different array filling fractions (Fig. 3): The required fraction of atoms are randomly removed.
  \item Vertical disorder (Fig. 2b): The harmonically trapped atoms are suddenly released and recaptured along the $z$ direction, preparing them in a Gaussian spatial distribution. The vertical lattice site of each atom is randomly sampled accordingly.
  \item Horizontal disorder: The horizontal atomic positions are randomly sampled from a continuous, uniform distribution within the area of the array.
\end{itemize}

% ++++++++++++++++++++++++++++++++++++++++

\subsection{Results}

The simulation results supporting the conclusions from the experimental measurements are presented in the following.

\subsubsection{Deviations from ideal array}

\begin{figure*}
  \centering
  \includegraphics{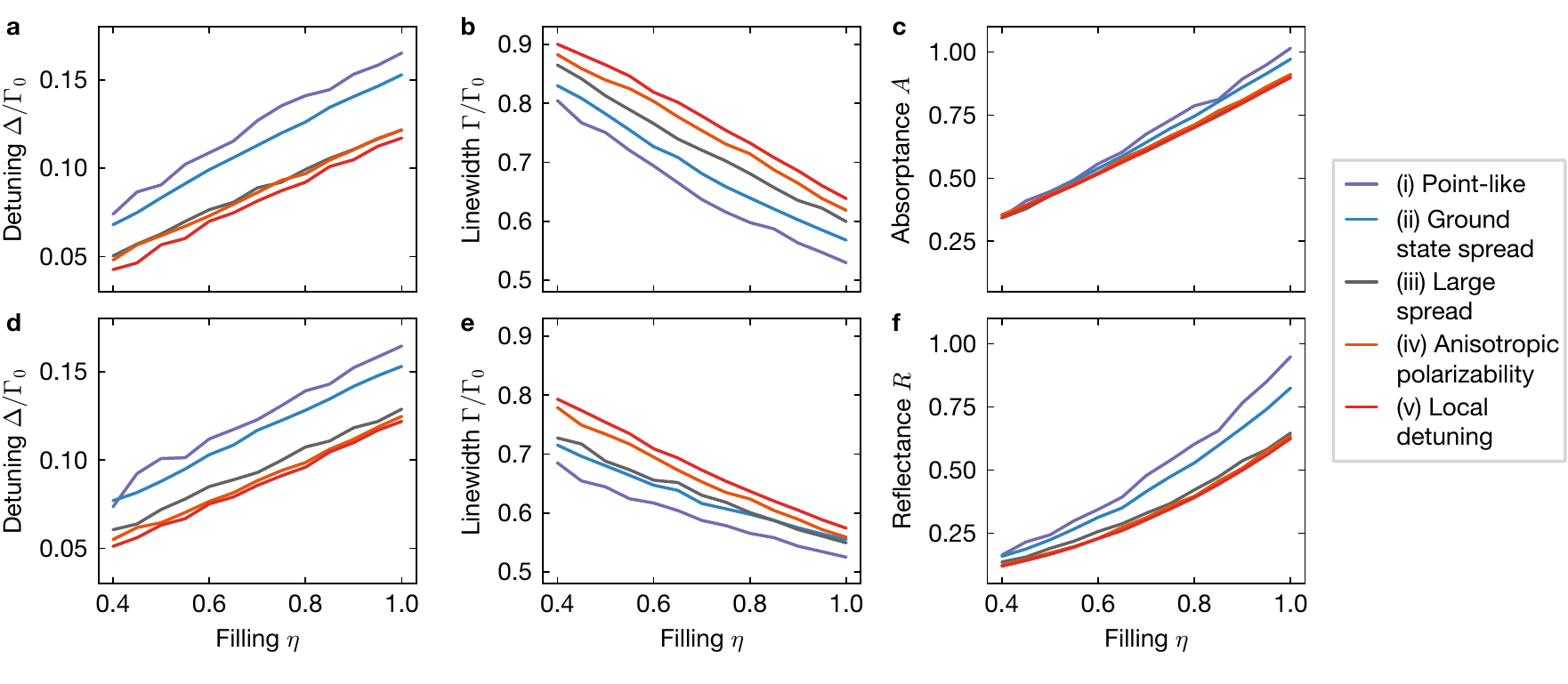}
  \caption{\textbf{Cooperative response of non-ideal arrays.}
    The top row of the figure shows the absorption simulations (a, b, c), the bottom row the reflection calculations (d, e, f). Both are obtained for the experimental numerical aperture of 0.68 and varying array filling fractions. The different graphs correspond to a perfect array with point-like, isotropically polarizable dipoles (i), a ground state wave function at lattice depth $V = 300~E_r$ leading to a positional spread of $\sigma_0 \simeq 0.054 a$ (ii), and a finite temperature state heated to a vertical positional spread of $\sigma_z \simeq 3 \sigma_0$ (iii). (iv) furthermore includes anisotropic polarizability, and (v) additionally takes into account spatially dependent resonance detunings.
}
  \label{fig:effects.filling}
\end{figure*}

Fig.~\ref{fig:effects.filling} shows how the cooperative electromagnetic response of an array at different filling fractions deviates from the isotropically polarizable, point-like response when various effects are taken into account.

As increasing the positional spread of the dipoles significantly modifies the directionality of the dipole emission field, it has a strong effect on the response amplitude, particularly reducing the reflectance (Fig.~\ref{fig:effects.filling}f), which is more sensitive than the absorptance (Fig.~\ref{fig:effects.filling}c) at a finite numerical aperture. Deviations from the ideal lattice structure also negatively affects the cooperativity, resulting in a reduced linewidth narrowing (Fig.~\ref{fig:effects.filling}b, e). The two depicted spreads refer to the shaded area shown in Fig. 3, and correspond to the ground state at a lattice depth of $V = 300~E_r$ and a finite temperature state, respectively. $E_r (\lambda) = h^2 / 2 m \lambda^2$ denotes the photon recoil energy, and the lattice depth is specified in terms of the lattice wavelength $1064\,\mathrm{nm}$. Due to dipole-dipole interactions, recoil heating in such collective systems is non-trivial \cite{robicheaux:2019}. As a rough estimate using a single-particle picture and assuming heating mainly along the $z$ axis, each scattering event deposits up to $4 E_r (780\,\mathrm{nm})$ of energy, so the plotted larger spread corresponds to more than $15$ scattered photons.

Taking into account the $\sigma^-$-anisotropy of the dipole polarizability allows us to probe the polarization of the scattered field. When the array is unity-filled, all dipole emission fields sum up to an in-plane polarization resembling the initial driving polarization. When the filling fraction is reduced, the polarization becomes less uniform (at $\eta = 0.4$, the average $\sigma^-$ fraction of the rescattered light is around $60 \%$), decreasing cooperativity. Consequently, the linewidth broadens faster, which can be observed from the larger slope in the reflectance linewidth (Fig.~\ref{fig:effects.filling}e).

As the local shifts in resonance are small compared to the natural linewidth, they only lead to a global inhomogeneous broadening of the linewidth (Fig.~\ref{fig:effects.filling}b, e). The overall influence on the resonance detuning is comparatively weak (Fig.~\ref{fig:effects.filling}a, d).

\subsubsection{Dependence on numerical aperture}

\begin{figure*}
  \centering
  \includegraphics{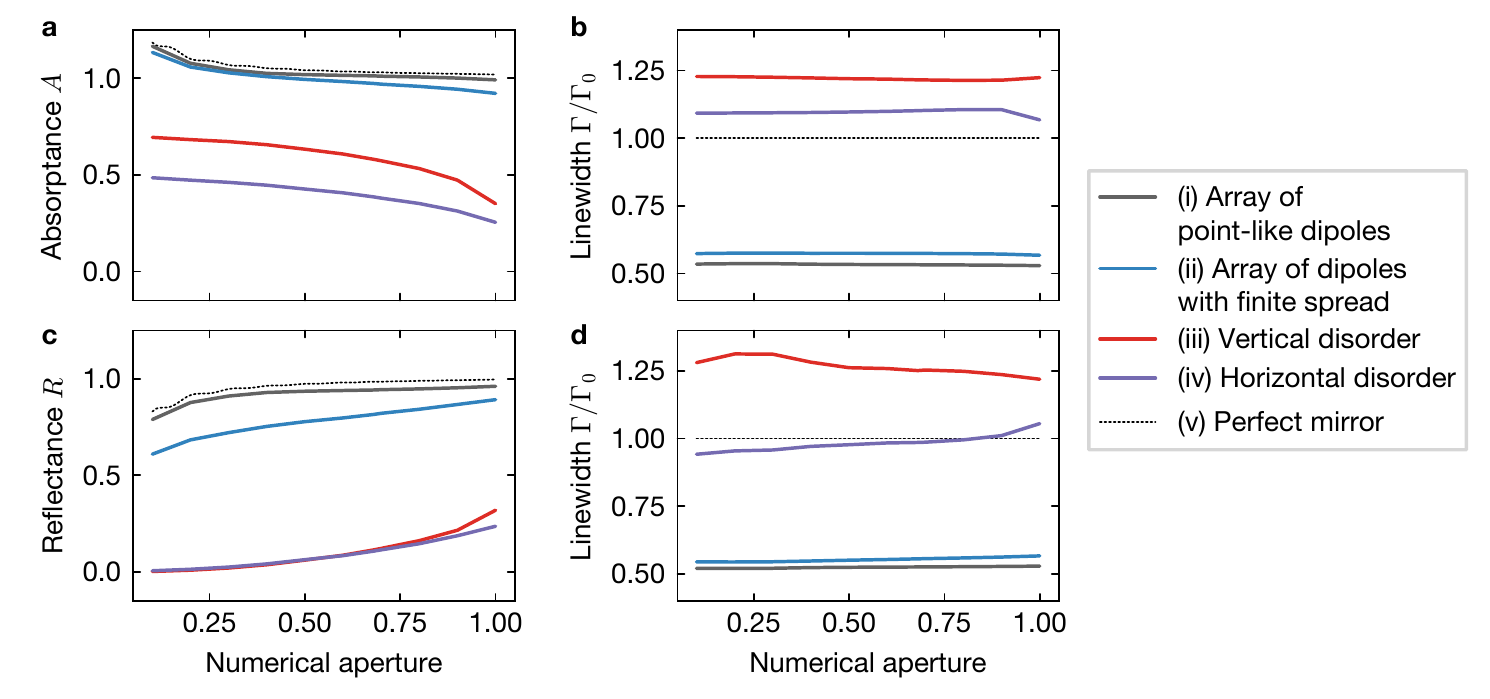}
  \caption{\textbf{NA-dependence of response for different spatial configurations.}
    The top row shows the simulated response in absorption (a, b) and the bottom row in reflection (c, d). The configurations include a $14 \times 14$ array of point-like dipoles (i), an array with positional spread corresponding to a lattice depth of $V = 300~E_r$ (ii), a vertically disordered array with a positional standard deviation of $\Delta z \simeq 10 a$ (iii), and horizontally disordered dipoles with the same size and density as the arrays (iv). The response amplitude is shown normalized by the filling fraction $\eta$. The dotted line depicts the response of a perfect mirror of the size of the array (v).
  }
  \label{fig:order.na_dependence}
\end{figure*}

\begin{figure*}
  \centering
  \includegraphics{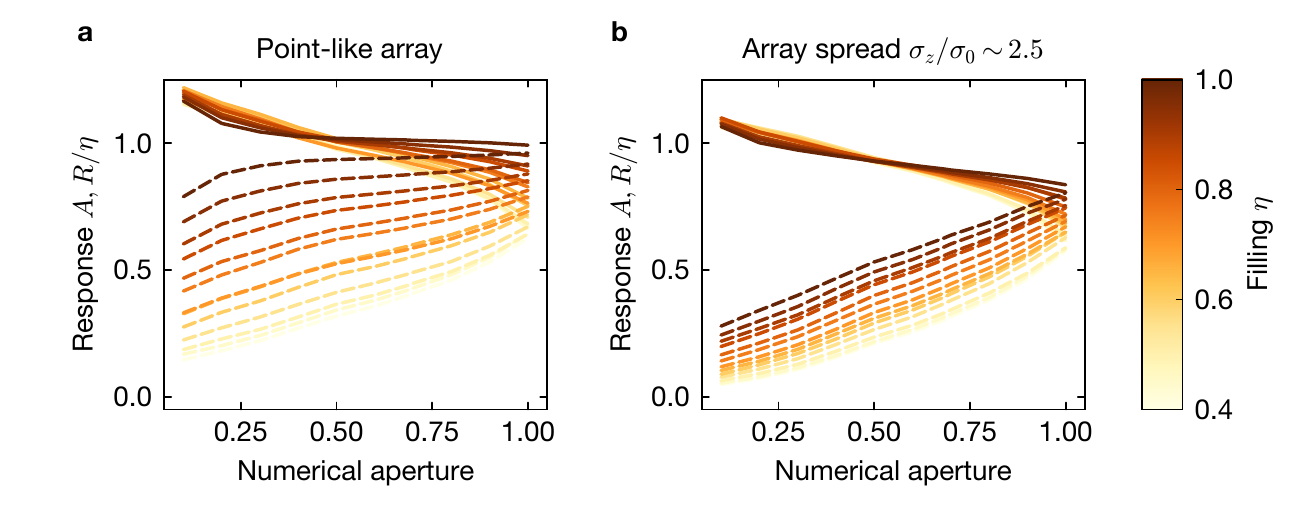}
  \caption{\textbf{Normalized array response amplitude for different fillings and spreads.}
    Simulated absorptance $A$ (solid) and reflectance $R$ (dashed) per atom of a $14 \times 14$ array consisting of point-like dipoles (\textbf{a}) and of dipoles with a finite horizontal spread of $\sigma_0 \simeq 0.054 a$ (\textbf{b}) at different array filling fractions $\eta$.
  }
  \label{fig:filling.na_dependence}
\end{figure*}

As the atomic array has a finite size, the scattered field forms a divergent emission pattern, such that the amount of collected light depends on the numerical aperture of the objective. Since unity-filled point-like arrays of dipoles are expected to have near-unity reflectance, these effects can be isolated by comparing their response to that of an actual mirror of the size of the array. Then deviations likely arise from diffuse scattering from the array. Here we consider a probe beam waist of $w_0 = 56 a$.

Fig.~\ref{fig:order.na_dependence} confirms the interpretation of the point-like array as a mirror, as it closely resembles the response of the mirror (Fig.~\ref{fig:order.na_dependence}a, c). The absorptance grows slightly above unity with decreasing NA as power is diffracted from the collimated beam into higher order modes which cannot be collected by the detector. Analogously, the reflectance decreases since less power is detected.

For a finite positional spread of the dipoles, specular reflectivity is still largely maintained, as the small slope at high NA indicates high directionality. This is not the case for both horizontally and vertically disordered configurations. In-plane disorder thus leads to highly divergent scattered fields, whereas similarities between vertical and horizontal disorder only in reflection suggest asymmetric divergence in the vertically disordered case. This can be understood by noting that the driving field imprints a corresponding phase onto the dipoles, such that the scattered field collimatedly interferes in the beam direction, but diverges in the opposite direction.

As diffuse scattering can be associated to exciting ensembles of different collective dipole modes, it is also of interest to study the NA dependence of the response linewidth. In the array configurations, where the primarily driven eigenmode is a directionally emitting subradiant state, the linewidth is rather NA-insensitive. The discrepancy between the disordered ensembles and the specularly reflecting arrays is also visible in the linewidth. Since the different modes have different linewidths and spatial emission patterns, the observed linewidth varies with NA.

Fig.~\ref{fig:filling.na_dependence} shows the NA-dependence of an array at varying filling fractions and differentiates between point-like dipoles and dipoles with positional spread. Energy conservation demands that at $\mathrm{NA} = 1$ absorptance must agree with reflectance, corresponding to the total scattered power. We indeed find this in our simulations, with small residual deviations resulting from numerical errors.

In the perfectly two-dimensional case of the point-like array (Fig.~\ref{fig:filling.na_dependence}a), the emission pattern is symmetric in forward and backward direction. As the probe beam is comparatively strongly collimated, this results in absorptance and reflectance curves which are symmetric with respect to the $\mathrm{NA} = 1$ value. When the filling fraction decreases, diffuse scattering and therefore the divergence increases, which can be seen from the steeper slopes of the curves. At the same time the cooperativity diminishes, which reduces the total scattered power (i.e. the $\mathrm{NA} = 1$ response amplitude). For finite numerical apertures both effects thus cancel in absorption but add up in reflection, resulting in the observed higher sensitivity in the reflection measurements.

Introducing a finite positional spread (Fig.~\ref{fig:filling.na_dependence}b) has a similar effect to adding disorder as in Fig.~\ref{fig:order.na_dependence}. The spread along the optical axis analogously leads to a discrepancy between the forward- and backward-emitted fields. Hence the slope of the curve becomes flatter in absorption but steeper in reflection, which again enhances the sensitivity of the reflection measurement.

\subsubsection{Linewidth broadening}

\begin{figure*}
  \centering
  \includegraphics{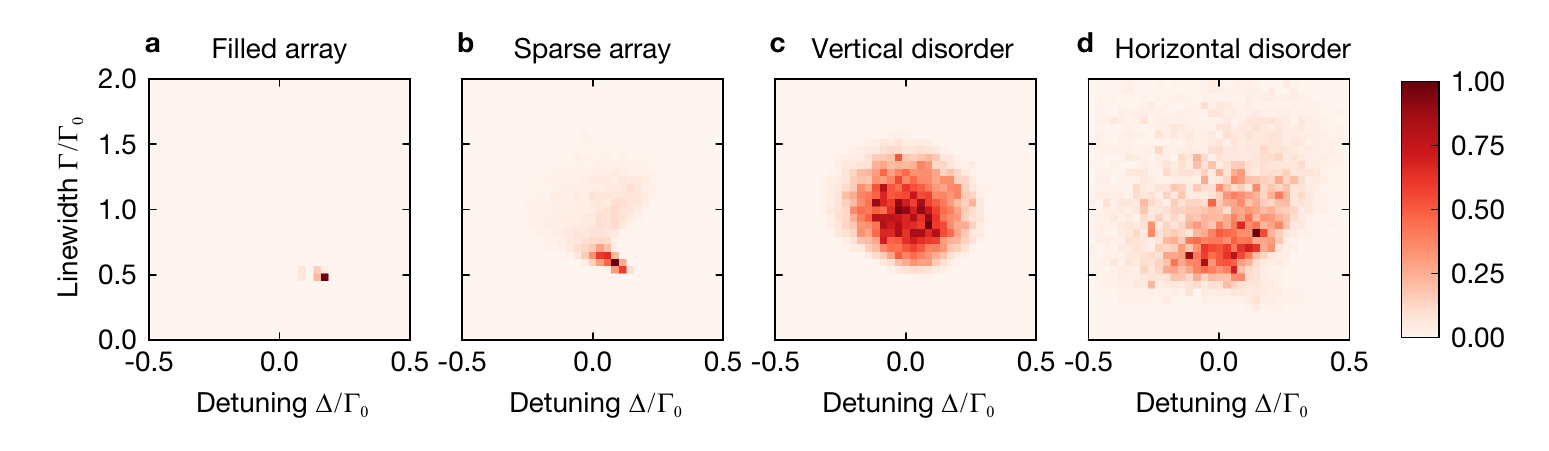}
  \caption{\textbf{Eigenmode decomposition of various spatial configurations for uniform drive.}
    The dipole eigenmode decompositions of a perfect array (\textbf{a}), an array with filling fraction $\eta = 0.4$ (\textbf{b}), a vertically disordered array with standard deviation $\Delta z \simeq 5 a$ (\textbf{c}), and an in-plane disordered dipole ensemble (\textbf{d}) are shown as mode amplitude probability densities. The 2D array configurations show excitation of mostly a single eigenmode, whereas disorder increases the number of modes. The substantial spread along the detuning axis for the disordered cases suggest significant inhomogeneous broadening effects as origin of increased measured linewidths.
  }
  \label{fig:mod.contr_pdf}
\end{figure*}

To determine the source of the different linewidths for the various spatial atomic configurations, it is useful to decompose the driven dipole patterns in their eigenmodes. The decomposition permits to distinguish whether a spectrally broadened response originates from inhomogeneous effects or from the excitation of superradiant modes.

Fig.~\ref{fig:mod.contr_pdf} compares the on-resonance dipole eigenmode decompositions for a point-like, unity-filled array, a sparsely filled array, an array with vertical positional disorder, and an in-plane randomized 2D ensemble.
It is evident that an eigenmode is only prepared for the perfect array (Fig.~\ref{fig:mod.contr_pdf}a). For a reduced filling fraction (Fig.~\ref{fig:mod.contr_pdf}b), effects from single-particle response become visible as mode contributions around the uncoupled response ($0, \Gamma_0$) start to grow. At large vertical spreads (Fig.~\ref{fig:mod.contr_pdf}c), the response consists of an ensemble of modes centred around the single-particle response, and whose deviations around this point decreases with increasing spread. The in-plane disordered system (Fig.~\ref{fig:mod.contr_pdf}d) has in comparison the largest mode spread, implying that the linewidth broadening partly results from superradiant modes but is still mainly caused by inhomogeneous effects.

\subsubsection{Probe beam size}

\begin{figure*}
  \centering
  \includegraphics{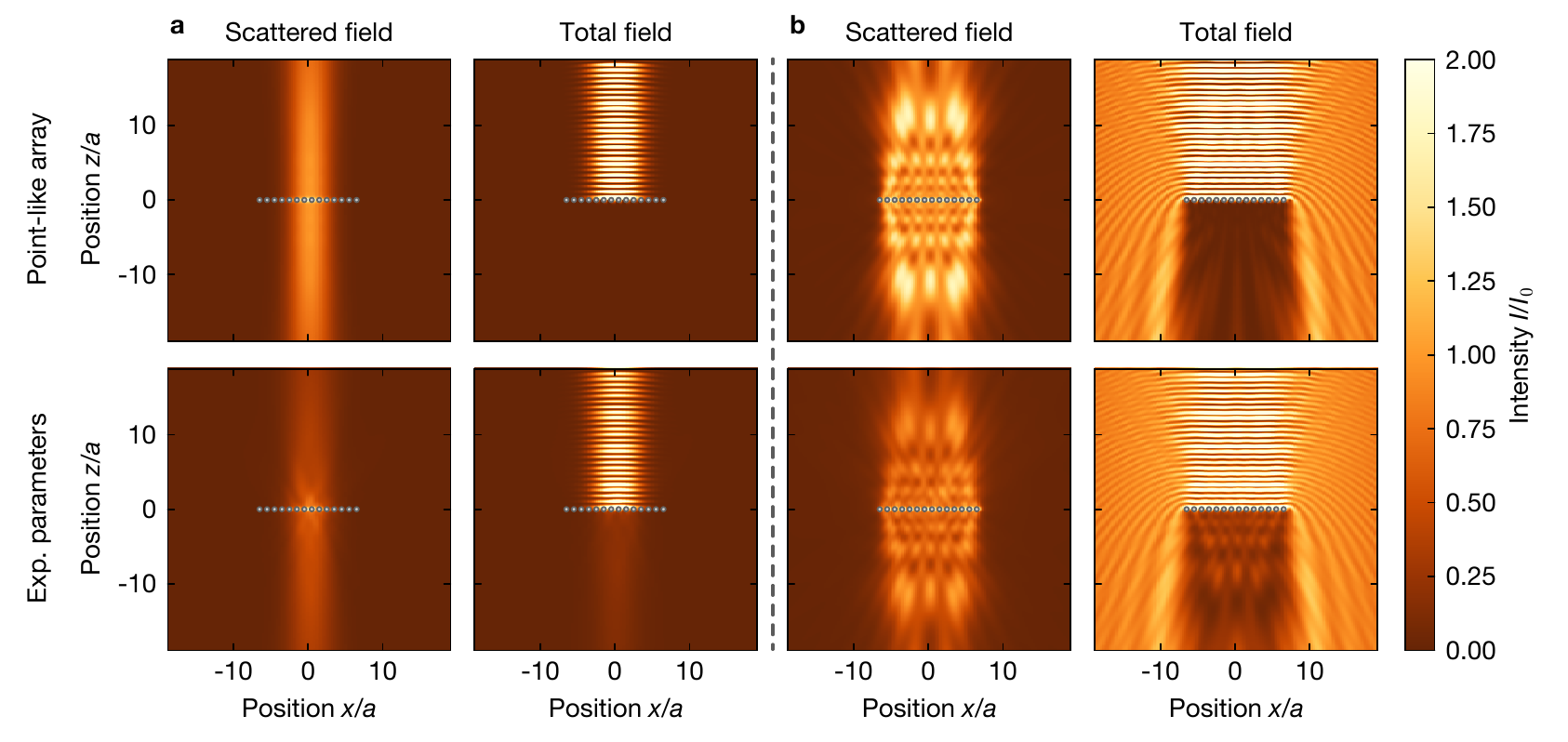}
  \caption{\textbf{Simulated intensity profiles showing mirror effect.}
    The probe beam in the calculations have waists of $w_0 = 4 a$ (\textbf{a}) and $w_0 = 56 a$ (\textbf{b}). The top row corresponds to a point-like dipole array, the bottom row to simulation parameters valid for our experimental configuration. The left column of each set of plots shows the emission pattern of the dipoles, the right ones display the superposition of the scattered with the incident field. When sending a beam smaller than the array, there is nearly perfect reflection. For larger beams, the imperfect mode matching leads to diffraction patterns. For a numerical aperture of 0.68, the reflectance for the large (small) beam at unity-filling is $R \simeq 0.95~(1.0)$ in the point-like dipoles case, and $R \simeq 0.8~(0.85)$ in the case of a ground state spread corresponding to a lattice depth of $V = 300~E_r$.
  }
  \label{fig:effects.propagation}
\end{figure*}

Since the probe beam used in the experiment is significantly larger than the area covered by the array, a full reflection of the total beam power cannot be observed. Additionally to the comparison with an ideal mirror in the previous sections, the response to a smaller beam can be simulated, which shows no significant deviation to the large beam case at our large numerical aperture.

The effect of a smaller beam can be directly seen in Fig.~\ref{fig:effects.propagation}, where the intensity profiles around point-like arrays and arrays realized in our experiment is shown.
For a beam smaller than the array size (Fig.~\ref{fig:effects.propagation}a), the point-like array is prepared in a dipole pattern, which perfectly cancels the incident field and thus forms a perfect mirror, which reflects the incident beam back into the same mode. For arrays deviating from the point-dipoles, the scattered field behaves qualitatively similar, but has an overall smaller extinction ratio.
When a larger beam is incident (Fig.~\ref{fig:effects.propagation}b), the ideal dipoles again perfectly cancel the incident field, forming a shadow which is only deteriorated due to edge diffraction effects. As the experimentally realized array responds similarly, analyzing a region of interest within the cloud size should thus probe similar physics as in the case of a small beam.

\clearpage
%%%%%%%%%%%%%%%%%%%%%%%%%%%%%%%%%%%%%%%%%%%%%%%%%%%
%                  Measurements                   %
%%%%%%%%%%%%%%%%%%%%%%%%%%%%%%%%%%%%%%%%%%%%%%%%%%%

\section{Additional measurements}

Here we present additional supporting material of the measurements to provide more details about the observations and conclusions in the main text.

\subsection{Natural linewidth reference}

\begin{figure}[b]
  \renewcommand\theHfigure{Supplement./thefigure}
  \centering
  \includegraphics{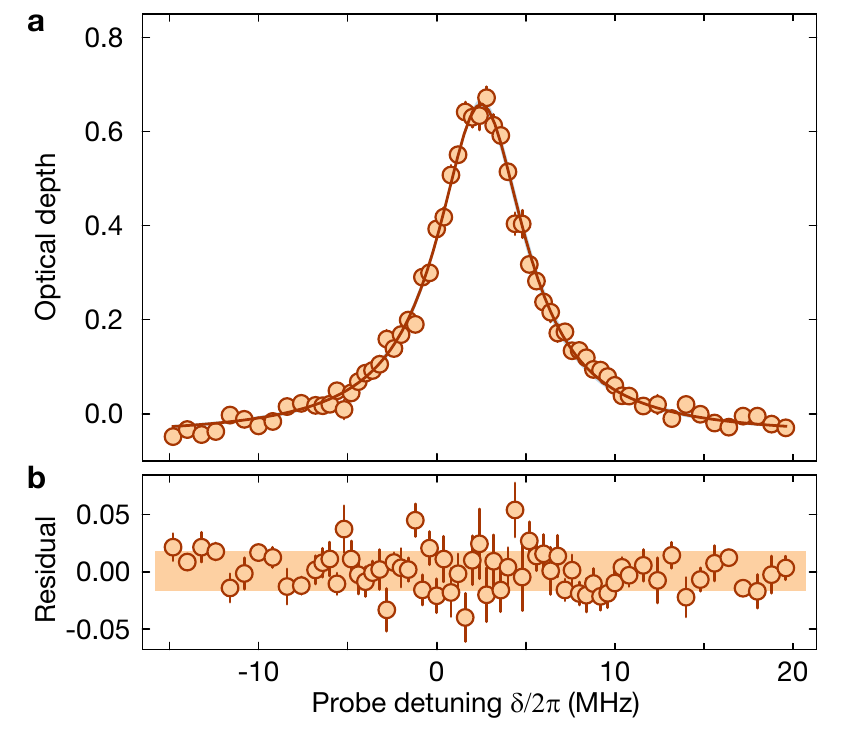}
  \caption{
  \textbf{Natural linewidth reference of a non-interacting 3d ensemble.}
    The dilute cloud is prepared by a free time-of-flight expansion after evaporation in a crossed optical trap. \textbf{(a)} The optical depth spectrum is fitted with a Lorentzian model, giving a linewidth of $\Gamma_\mathrm{ref} = 2 \pi \times 6.04(11)$ MHz close to the natural linewidth of $\Gamma_0 = 2 \pi \times 6.067$ MHz for single atoms. \textbf{(b)} Fitting residuals of the measured spectrum. Error bars denote s.e.m.
  }
  \label{fig.SI_E1}
\end{figure}

To compare our cooperative linewidth measurements with the optical transition linewidth of ``non-interacting'' atoms, we prepared a dilute three-dimensional (3d) cloud and probed the absorption with the $\ket{F = 2} \rightarrow \ket{F' = 3}$ transition in the $\mathrm{D}_2$-line manifold of $^{87}$Rb, similar to the main text. To prepare the cloud, we first evaporated the atoms, in the $\ket{F = 2, m_F = -2}$ state, within a crossed optical dipole trap (without loading into the optical lattices), and then switched off the optical traps for 10\,ms to dilute the atomic density before the onset of Bose-Einstein condensation. After that, we performed absorption imaging of the 3d cloud with the same ULE cavity-locked probe beam as in the main text. To avoid broadening the optical transitions, we compensated the ambient magnetic field to be less than $10\,\mathrm{mG}$, making the Zeeman sublevels degenerate both in the ground and excited states.

The resulting optical density spectrum $\mathrm{OD} = -\log T$ is shown in Fig.~\ref{fig.SI_E1} and yields a fitted Lorentzian linewidth of $\Gamma_\mathrm{ref} = 6.04(11)\,\mathrm{MHz}$, in agreement with the natural linewidth $\Gamma_0 = 6.067(2)\,\mathrm{MHz}$ \cite{steck_rb:2019} of a single atom. The observed linewidth confirms that the broadening due to the laser linewidth in the probe beam can be safely neglected in the analysis.

% ++++++++++++++++++++++++++++++++++++++++

\subsection{Calibration of intensity in the reflection probe}

\begin{figure*}
  \centering
  \includegraphics{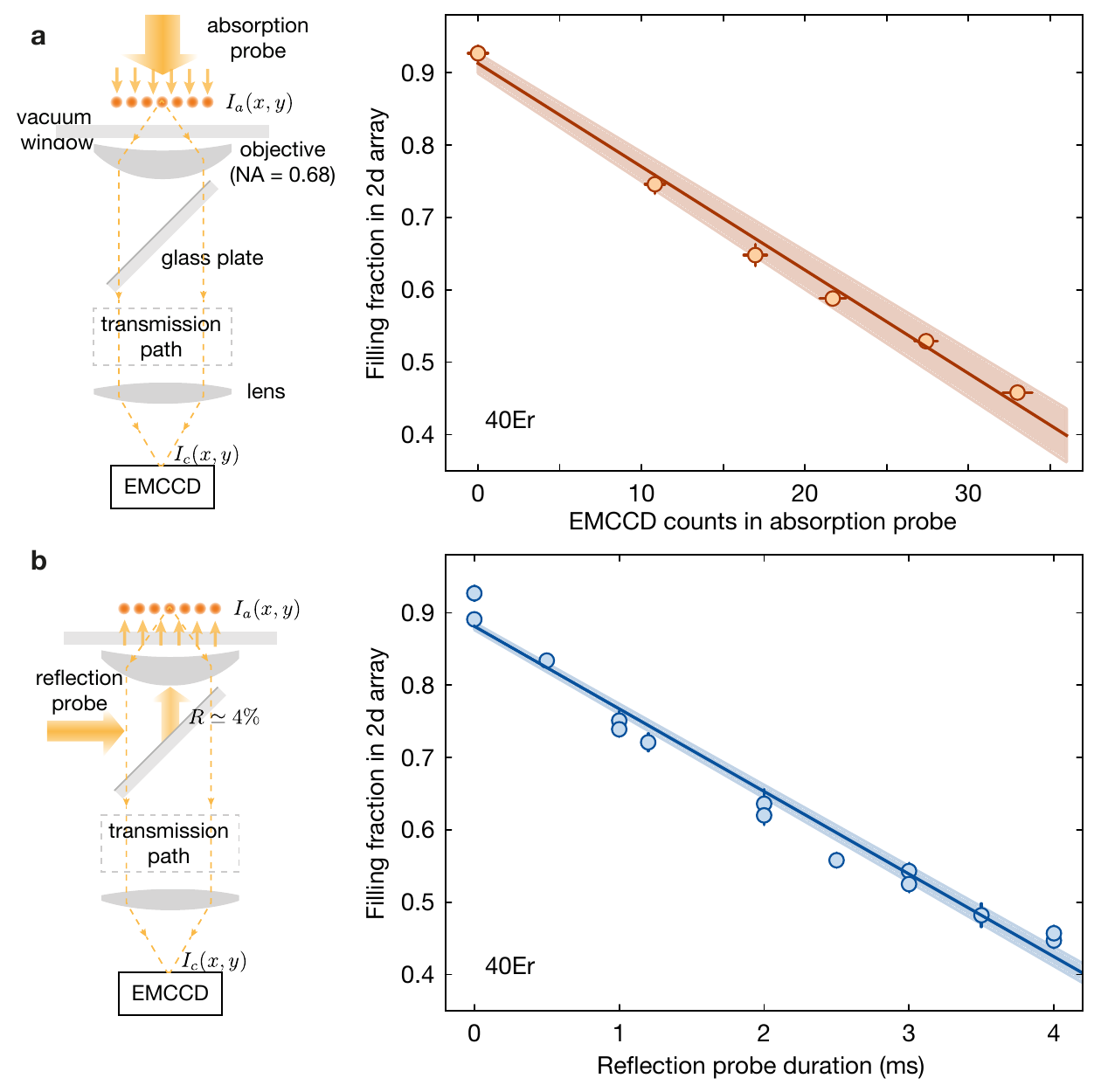}
  \caption{\textbf{Intensity calibration of the reflection probe.}
    \textbf{(a)} Lattice filling versus resonant heating by the transmission probe ($V_{x,y,z}=40\,E_r$). The light field at the position of the atoms is imaged onto the EMCCD, and the photon count in each binned pixel is recorded. With the same resonant heating, the filling of the atoms in the 2d array is independently characterized with single-site resolved fluorescence imaging. The measured lattice filling versus photon counts, with linear fit, is shown at the right side.
    \textbf{(b)} Lattice filling versus resonant heating by the reflection probe ($V_{x,y,z}=40\,E_r$). A small fraction ($\approx 4\%$) of the reflection probe is reflected by a glass plate and goes through the objective onto the atoms. The measured lattice filling versus the duration of the reflection probe, with linear fit, is shown at the right side.
    Shaded regions correspond to one s.d. of best fitted parameters. Error bars denote s.e.m.
  }
  \label{fig.SI_E2}
\end{figure*}

For the absorption measurement in the main text, the probe field can be directly recorded by the EMCCD through the same optical path as the transmitted field, from which the transmittance through the atom array can be evaluated, as shown in Fig.~\ref{fig.SI_E2}a. However, the intensity of the reflection probe cannot be directly measured by the EMCCD, as it is sent opposite to the imaging path as shown in Fig.~\ref{fig.SI_E2}b. We utilized a lattice thermometry to estimate the photon count in the reflection probe beam, based on atom loss after variable resonant optical pushing in a shallow lattice potential. 

In the experiment, we prepared a nearly unity filled 2d array of single atoms as described in the main text and then held them with lattice depths of $V_{x,y,z}=40~E_r$ for all three lattices. Afterwards, we shone a resonant probe beam for a specific duration onto the atoms, and then checked the filling $\eta$ of the remaining atoms in the optical lattice using single-site resolved fluorescence imaging \cite{sherson:2010}. In the first measurement, we calibrated the dependence of the lattice filling on the intensity of the transmission probe as shown in Fig.~\ref{fig.SI_E2}a, in which the measured photon counts $I^{\mathrm{a}}_{\mathrm{px}}$ per binned pixel on the EMCCD is plotted in the horizontal axis. From this measurement, we observed that a linear fit model described our data, with a fitted slope of $d\eta / dI^{\mathrm{a}}_{\mathrm{px}} = 0.0143(6)$. After that, we turned to measure the dependence of filling on the duration $T$ of the reflection probe, with the measured values shown in Fig.~\ref{fig.SI_E2}b. A linear fitting gives a coefficient of $d\eta / dT = 0.114(2)$ per ms. For both beams, we confirmed that the beam sizes are significantly larger than the size of the atom array, such that we could treat the intensities of both beams as homogeneous in the spatial region of the atoms. The comparison between the two coefficients yields an estimated reflection probe beam photon flux of $I^{\mathrm{r}}_{\mathrm{px}} = 8.0(4)$ photon counts per ms per binned pixel.

We used the above calibrated quantity to characterize the photon counts in the reflection probe at the position of the atoms throughout this work. As the intensity pattern recorded by the EMCCD is a direct image of the local field in the plane of the atoms, the reflectance of the array can be given as the measured reflected photon counts from the reflection probe divided by the above calibrated photon counts in the reflection probe.

% ++++++++++++++++++++++++++++++++++++++++

\subsection{Response of a 2d pancake of randomly positioned atoms}

\begin{figure}
  \centering
  \includegraphics[width=0.95\columnwidth]{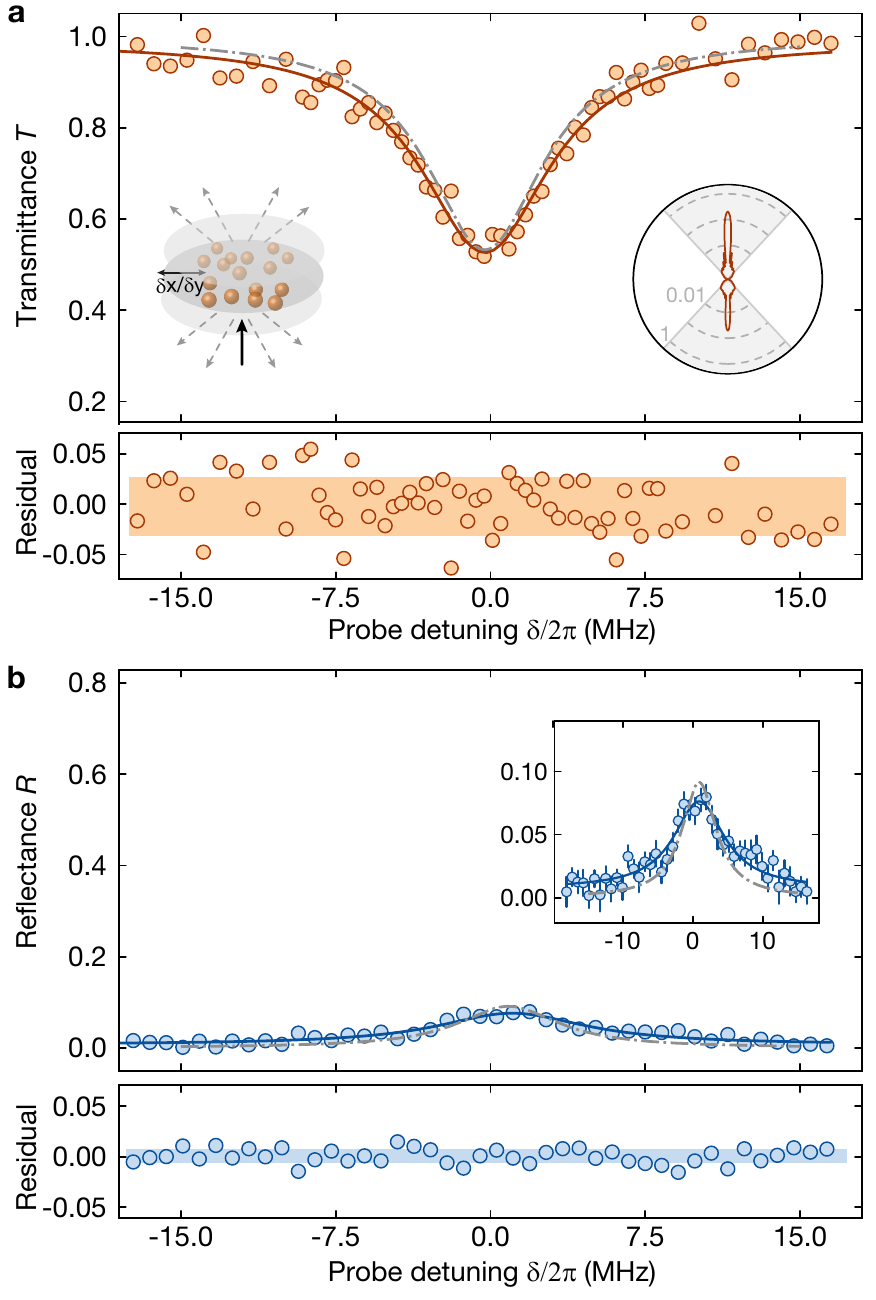} 
  \caption{\textbf{Cooperative response of disordered atoms in a 2d pancake.}
 The linewidths are significantly broadened to be $\Gamma = 2 \pi \times 7.63(0.34)$\,MHz and $\Gamma = 2 \pi \times 8.8(13)$\,MHz in the absorption and reflection spectra, with a transmittance of $T = 0.54(1)$ and a considerably smaller reflectance of $R = 0.07(1)$, respectively. The dashed-dotted lines are numerically simulated spectra, a mean density of $0.9$ ($0.6$) is used for transmission (reflection). Inset of $\textbf{(a)}$, sketch of the ordering of the atoms in a single 2d pancake, and the differential scattering cross-section with a mean density of 0.8 obtained by numerical simulation.
 Error bars denote s.e.m.
  }
  \label{fig.SI_E3}
\end{figure}

In addition to the two spatial configurations compared in Fig.~2 of the main text and disordered ensembles studied in \cite{pellergrino:2014, corman:2017}, we also probed another geometry of the atoms, in which they are confined in a single 2d pancake but with their horizontal positions completely disordered. To prepare such a disordered 2d pancake, we started with a nearly unity filled 2d array at $V_{x,y}=40\, E_r$ in the horizontal lattices and $V_{z}=300\,E_r$ in the $z$-lattice. Then we suddenly switched off the horizontal lattices in less than $100\,\mu$s while keeping the vertical lattice unchanged. The atoms immediately start randomly distributing within the 2d pancake potential. In this way, the 2d cloud with continuous position disorder has a similar average density as the 2d ordered array during illumination. To avoid expansion and thus diluting the atom ensemble, we reduced the duration of the transmission probe to $200\,\mu$s and that of the reflection probe to $2.3$\,ms, with a total photon number of $\approx 20(50)$ per lattice site for absorption (reflection), similar to the other measurements. The measured absorption and reflection spectra are shown in Fig.~\ref{fig.SI_E3}. A maximal transmission of $T=0.54(1)$ and a linewidth of $\Gamma/\Gamma_0=1.26(6)$ was observed in the transmission signal, and a much lower reflectance of $R=0.07(1)$ together with a broadened linewidth of $\Gamma/\Gamma_0=1.5(2)$ in reflection. While the measured transmission spectrum is quite close to the simulated curve, we note that there can be a significant deviation between the reflection spectrum and the simulated curve with a mean density of $0.9$ (here the simulation of reflection uses a mean density of $0.6$). We attribute this to the limited control over the horizontal motion of the atoms during optical probing, as on resonance we observed a significant increase in the cloud size in reflection. This effect is much weaker in the absorption measurement due to the smaller photon number and shorter probe duration.

% ++++++++++++++++++++++++++++++++++++++++

\subsection{Response versus analysis region-of-interest}

\begin{figure*}
  \centering
  \includegraphics{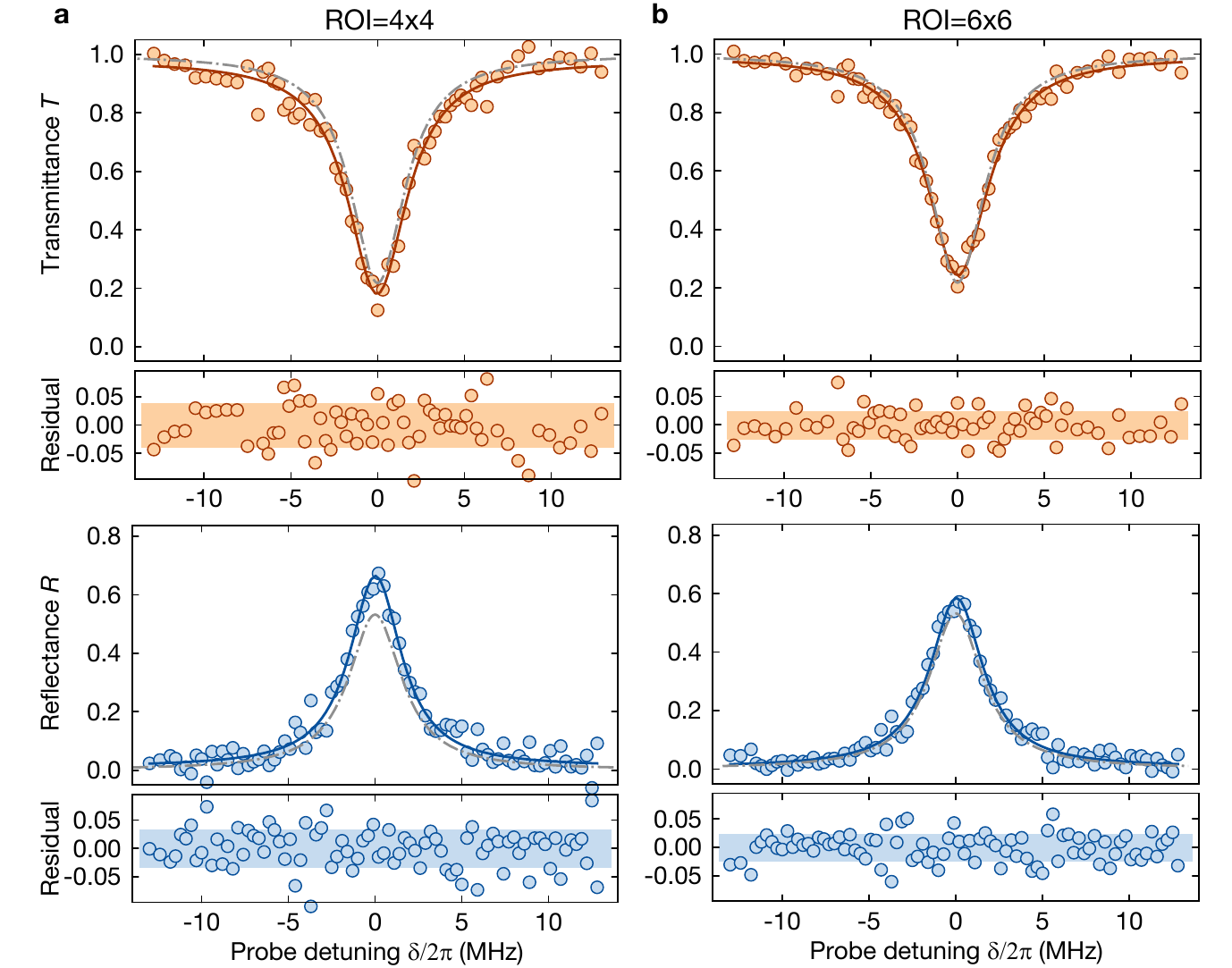} 
  \caption{\textbf{Response dependence on analysis region-of-interest.} 
  \textbf{(a)} Spectrum with an analysis ROI of $4\times4$ binned pixels. In the transmission spectrum, the fitted absorptance is $A = 0.80(1)$, with a linewidth of $\Gamma = 2 \pi \times 4.15(14)$ MHz. For the reflection measurement, the fitted reflectance is $R = 0.65(5)$, with a linewidth of $\Gamma = 2 \pi \times 3.63(14)$ MHz.
  \textbf{(b)} Spectrum with an analysis ROI of $6\times6$ binned pixel. In the transmission spectrum, the fitted absorptance is $A = 0.75(1)$, with a linewidth of $\Gamma = 2 \pi \times 4.22(11)$ MHz. For the reflection measurement, the fitted reflectance is $R = 0.58(5)$, with a linewidth of $\Gamma = 2 \pi \times 3.71(12)$ MHz.
  The residuals of Lorentzian fit is also shown together, indicating that the Lorentzian model works well for the spectral analysis.
   Error bars denote s.e.m.
  }
  \label{fig.SI_E4}
\end{figure*}

Due to the limited size of the array, we could always observe a slightly higher absorptance or reflectance at the center of the array. However, in order to improve the signal-to-noise ratio, we analyzed all spectra with a ROI of $6\times6$ binned pixels, covering $\simeq126$ lattice sites, in the main text. Here we show the spectra with a smaller ROI of $4\times4$, covering $56$ atoms in the array, for a nearly unity filled 2d array as shown in Fig.~\ref{fig.SI_E4}a. As a comparison, we also present the same spectrum with a larger ROI of $6\times6$ in Fig.~\ref{fig.SI_E4}b. The fitted linewidths are very close to each other for the two ROI sizes, in both transmission and reflection spectrum. However, we indeed observe that the absorptance is improved from $0.75(1)$ to $0.80(1)$, and the reflectance is increased from $0.58(5)$ to $0.65(5)$, when the ROI is reduced from $6 \times 6$ to $4 \times 4$ binned pixels. With the residuals of the Lorentzian fitting also shown in the plots, it shows that such a fitting works well for the spectral analysis.

% ++++++++++++++++++++++++++++++++++++++++

\subsection{Response versus photon recoil heating}

\begin{figure*}
  \centering
  \includegraphics{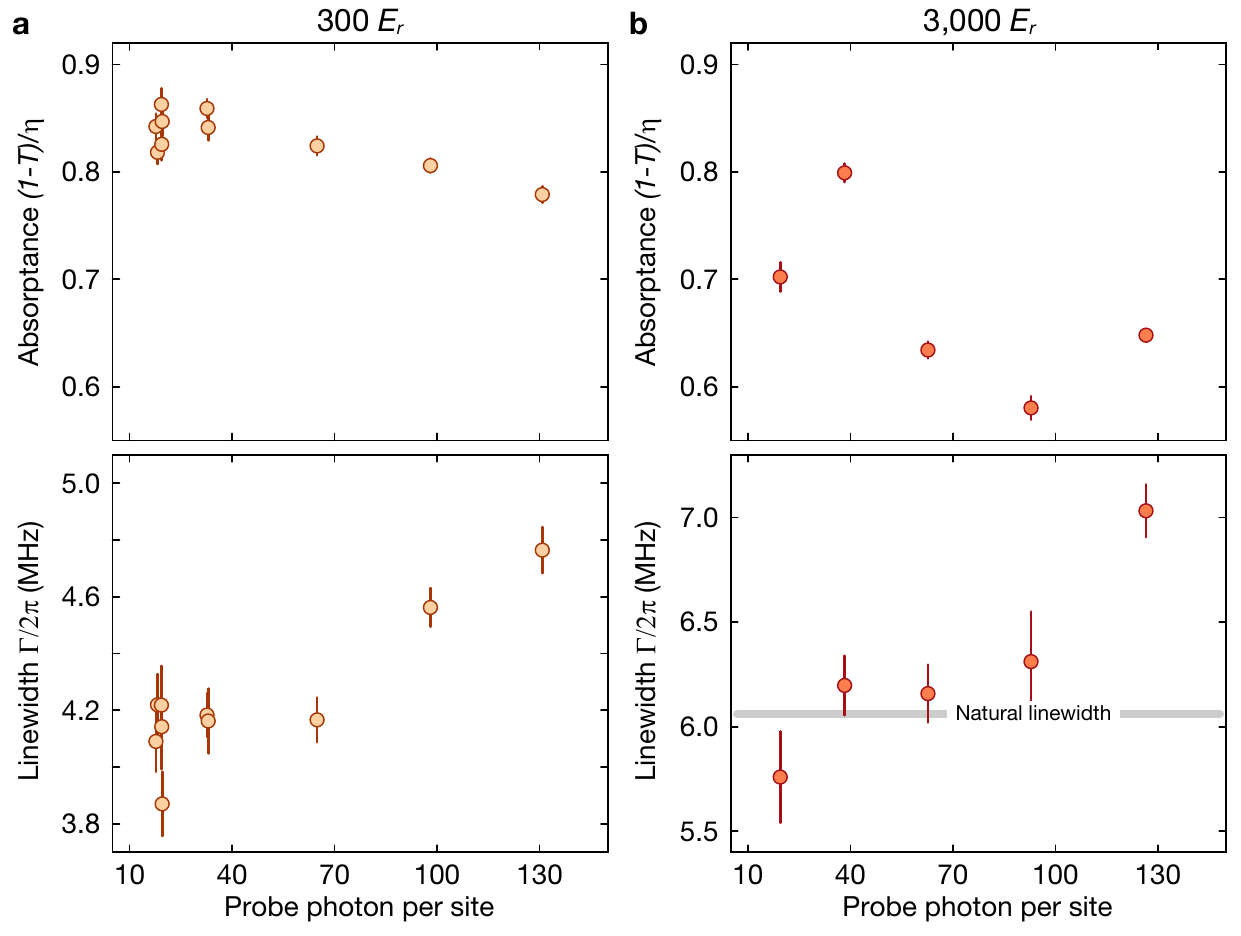} 
  \caption{\textbf{Effects of photon recoil heating in the cooperative response.}
    \textbf{(a)} Measured absorptance and linewidth for different probe photon counts at $V_{x,y,z}=300\,E_r$ for all three lattices. 
    \textbf{(b)} Measured absorptance and linewidth for different probe photon counts at $V_{z}=3,000\,E_r$ for the vertical lattice, and $V_{x,y}=1,000\, E_r$ for the two horizontal lattices. 
    The horizontal axis denotes the photon number per lattice site converted from the photon counts recorded by the EMCCD.
    Error bars denote s.d. of the fitted values.
  }
  \label{fig.SI_E5}
\end{figure*}

In the measurements of the main text, we typically used $\simeq 35$ photon counts for each binned pixel and pulse duration of $3$ ms in the transmission probe, corresponding to $\simeq 20$ incoming photons per lattice site and a flux of $\simeq 7$ photons per ms for each lattice site. For the reflection probe, the estimated photon count is $45/85$ for each binned pixel, corresponding to $27/50$ photons in total per lattice site and a flux of $\simeq 5$ photons per ms for each lattice site. For all of the above probing intensities, we have not observed any loss of atoms in the $V_{x,y,z}=300~E_r$ deep optical lattices after the resonant optical probing during the measurements. 

To check the dependence of the cooperative response of the atom array on recoil heating, we measured the absorption under different probe photon counts for two different lattice depths. In the first measurement, all three lattice depths were kept at $V_{x,y,z}=300\,E_r$ and the measured absorptance and linewidth is shown in Fig.~\ref{fig.SI_E5}a. While the absorptance does not clearly change, the fitted linewidth starts increasing until the probe beam goes beyond $70$ photons per lattice site. In the second measurement, the horizontal lattices were kept at $V_{x,y}=1,000\,E_r$ but the vertical lattice was prepared at $V_{z}=3000\,E_r$. In this case, there is an overall reduction in the absorptance compared with the lower lattice case, and the linewidth reaches up to 7 MHz for the high photon counts, as shown in Fig.~\ref{fig.SI_E5}b.

% ++++++++++++++++++++++++++++++++++++++++

\subsection{Response versus photon flux}

\begin{figure}
  \centering
  \includegraphics[width=0.95\columnwidth]{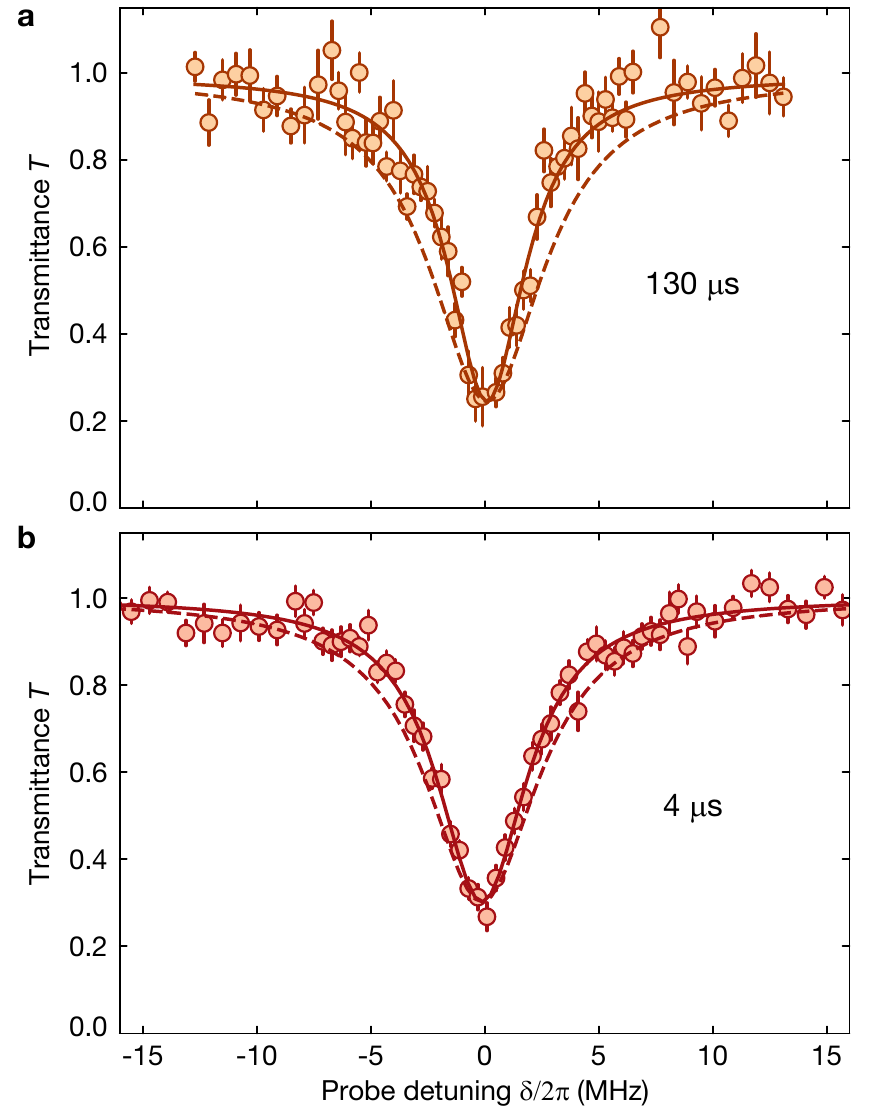}
  \caption{\textbf{Effects of different fluxes with the same total photon counts.}
    \textbf{(a)} Measured transmission spectrum for a probe pulse width of $130~\mu \mathrm{s}$, giving an absorptance of $0.75(2)$ and linewidth of $4.08(21)$ MHz.
    \textbf{(b)} Measured transmission spectrum for a probe pulse width of $4~\mu \mathrm{s}$, giving an absorptance of $0.70(1)$ and linewidth of $4.80(18)$ MHz.
    Both measurements have the same photon counts in the probe beam. Solid line shows the best fit with a Lorentzian model. Dashed lines represent the reference spectrum with the fitted linewidth replaced by the natural linewidth. Error bars in the plots denote s.e.m.
  }
  \label{fig.SI_E6}
\end{figure}

Beside the photon counts, we also compared the cooperative response of the nearly unity filled atom array under different photon fluxes, with the same total photon number of $\simeq 20$ per lattice site. For the absorption measurements in the main text, the transmission probe beam has a flux of $\simeq 7$ photons per ms for each lattice site, which corresponds to $1.8 \times 10^{-4}$ incident photons per lifetime of a single atom. In Fig.~\ref{fig.SI_E6}a, we increased the probe intensity to $4 \times 10^{-3}$ photons per lifetime of a single atom, with a total duration of $130\,\mu$s in the probe beam. The measured absorptance is $A = 0.75(2)$ and the linewidth is $\Gamma = 2 \pi \times 4.08(21)$, which is similar to the values in the main text. In Fig.~\ref{fig.SI_E6}b, the photon flux is further increased to $0.14$ photons per lifetime of a single atom, with a total duration of $4~\mu$s in the probe beam. The measured absorptance is slightly reduced to $A = 0.70(1)$ and the linewidth is broadened to $\Gamma = 2 \pi \times 4.80(18)$ MHz. We note that with the flux used here, more than one collective excitations would be simultaneously created in the array, but the cooperative response is still safely in the subradiant regime. Besides, Fourier broadening in the probe beam due to the short pulse significantly contributes to the larger linewidth observed. Therefore it is hard to conclude any further interaction effects in the multiple excitation regime with the current measurement.

\bigskip
% ++++++++++++++++++++++++++++++++++++++++

\subsection{Light shifts of the anti-trapped excited state}

\begin{figure}
  \centering
  \includegraphics[width=0.95\columnwidth]{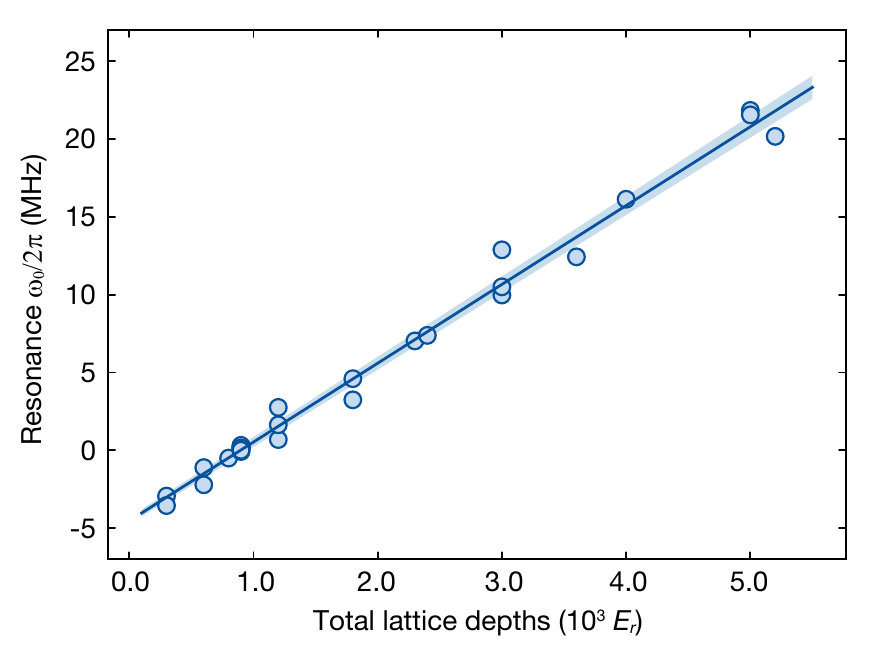}
  \caption{\textbf{Light shift of the optical transition frequency.}
    The fitted resonance frequency versus the total lattice depth is shown, giving a trapping parameter of $5.06(10)~\mathrm{kHz} / E_r$. Excluding the trapping depth of the ground state, we can derive the trapping coefficient for the excited state to be $3.03(10)~\mathrm{kHz} / E_r$. Shaded region corresponds to the fitted parameters with one s.d. Error bars are smaller than the circles of the data points.
  }
  \label{fig.SI_E7}
\end{figure}

As the excited state $5^{2}P_{3/2}$ in the probed optical transition is anti-trapped in the $1064$\,nm wavelength optical lattice, the transition frequency depends on the actual lattice depth used in the measurement. With the lattice dependence measurement shown in the main text, we can also identify the anti-trapping parameters of the $\ket{5 ^{2}P_{3/2}, F'=3, m_F'=-3}$ excited state in the lattice potential. The fitted resonance frequencies under different total lattice depths is shown in Fig.~\ref{fig.SI_E7}, from which the transition frequency is fitted to have a dependence of $5.06(10)~\mathrm{kHz} / E_r$ in the total lattice depth. As the ground state has a potential dependence of $2.03~\mathrm{kHz}/E_r$, the light shift of the excited state in the lattice potential is $3.03(10)~\mathrm{kHz} / E_r$.

\clearpage

\end{document}